\documentclass[%
 reprint,showpacs,prd,
 amsmath,amssymb,
 aps,
prd,
]{revtex4-1}

\usepackage{graphicx}
\usepackage{dcolumn}
\usepackage{multirow,color}
\usepackage{bm}

\begin{document}


\title{Quarkyonic Matter Equation of State in Beta-Equilibrium}
\author{Tianqi Zhao }
\affiliation{Dept. of Physics \& Astronomy, Stony Brook University, Stony Brook, NY 11794-3800}
\email{tianqi.zhao@stonybrook.edu}
\author{James M. Lattimer}
\affiliation{Dept. of Physics \& Astronomy, Stony Brook University, Stony Brook, NY 11794-3800}
\email{james.lattimer@stonybrook.edu}
\date{January 2020}

\begin{abstract}
Quark matter may appear due to a hadronic-quark transition in the core of a  hybrid star. Quarkyonic matter is an approach in which both quarks and nucleons appear as quasi-particles in a crossover transition, and provides an explicit realization of early ideas concerning quark matter (e.g., the MIT bag model).  This description has recently been employed by McLerran and Reddy to model chargeless (pure neutron) matter with an approach that has the virtue that the speed of sound rises quickly at a neutron-quark transition so as to satisfy observational constraints on the neutron star maximum mass ($\gtrsim2M_\odot$) and the radius of a $1.4M_\odot$ star ($R_{1.4}\lesssim 13.5$ km).  Traditional models involving first-order transitions result in softer pressure-energy density relations that have difficulty satisfying these constraints except with very narrow choices of parameters.   We propose a variation of quarkyonic matter involving protons and leptons whose energy can be explicitly minimized to achieve both chemical and beta equilibrium, which cannot be done in the chargeless formulation.  Quarkyonic stellar models are able to satisfy observed mass and radius constraints with a wide range of model parameters, avoiding the obligatory fine-tuning of conventional hybrid star models, including requiring the transition density to be very close to the nuclear saturation density.  
Our formulation fits experimental and theoretical properties of the nuclear symmetry energy and pure neutron matter, and contains as few as three free parameters.  This makes it an ideal tool for the study of high-density matter that is an efficient alternative to piecewise polytrope or spectral decomposition methods. \end{abstract}

\maketitle


\section{Introduction}
The measurement of neutron stars masses (see \citep{lattimer2012nuclear, ozel2016masses} for general reviews) greater than or equal to $2M_\odot$ has proven to be a powerful constraint on the dense matter equation of state.  Equally important have been advances in gravitational wave measurements of the binary neutron star merger GW170817 \cite{abbott2017gw170817, de2018tidal, abbott2018gw170817, capano2019gw170817} and NICER X-ray observations of PSR J0030+0451 \cite{riley19,miller2019psr} that, combined, imply that $1.4M_\odot$ stars have radii $R_{1.4}\lesssim 13.5$ km \cite{raaijmakers2019constraining}.  The latter result is supported by nuclear symmetry energy measurements \cite{tsang2009constraints,abrahamyan2012measurement,zhang2013constraining,rossi2013measurement,danielewicz2014symmetry,roca2013electric,lynch2018nuclear,lattimer2013constraining} and theoretical chiral Lagrangian calculations of pure neutron matter\cite{gandolfi2012maximum}.  These constraints, coupled with causality considerations, strongly imply that the equation of state (EOS) of dense matter quickly changes from softness in the density range 1-2$n_s$, where $n_s\simeq0.16$ fm$^{-3}$ is the nuclear saturation density, to relative stiffness at higher densities, with sound speeds approaching the speed of light.  At yet higher densities, in the vicinity of the maximum densities found in neutron stars (5-10$n_s$) and beyond, matter is expected to consist of deconfined quark matter with sound speeds approaching the relativistic value $\sqrt{1/3}c$. {Many f}requently-used parameterizations of the dense-matter EOS, including power series expansions in nucleon wavenumber $k$ or baryon density $n$ \cite{tews2017symmetry}, piecewise polytropes \cite{read2009constraints}, constant-sound speed segments \cite{zdunik2013maximum,han2019tidal}, spectral decomposition \cite{lindblom2010spectral}, and relativistic mean-field theoretical approaches \cite{hornick2018relativistic}, fail to capture these trends.  In addition, the transition between hadronic matter and quark matter, if first-order, introduces significant softening \citep{alford2005hybrid,han2019tidal,chatziioannou2020studying} in the EOS precisely in the region where it must become stiff to satisfy the observational and experimental constraints, suggesting a crossover transition \cite{baym2018hadrons,baym2019new,han2019treating} instead.  

{Recently, to satisfy these assorted criteria, speed of sound parameterizations \cite{tews2018critical,greif2019equation} have been introduced to introduce a large increase in the speed of sound at intermediate densities followed by a decrease so that the expected asymptotic $\sqrt{1/3}c$ behavior ensues.  These approaches explicitly modeled the sound speed with arbitrary Gaussian-like functions.  However,
 an alternative physically-intuitive theoretical} approach, the quarkyonic model, has been proposed \citep{mclerran2019quarkyonic}.
In the quarkyonic model, at low densities, quarks are confined within nucleons interacting via a conventional potential.  However, when the nucleon momenta surpass critical values at the transition density $n_t$ between nuclear and quarkyonic matter, the low momenta degrees of freedom inside the Fermi sea are treated as non-interacting quarks, and at higher momenta they are treated with confining forces resulting in baryons. In this model, nucleons are viewed as triplets of quarks near the Fermi surface, while the free quarks resemble the  massive quarks having a dressed gluon cloud of the constituent quark model. 

The appearance of free quarks above the quarkyonic transition gives nucleons extra kinetic energy by pushing them to higher average momenta while decreasing their densities. In a sense, the quarks 'drip out' of the nucleons and fill the lowest momentum states, resulting in a rapid increase in pressure.  These considerations culminate in a dramatic increase in sound speed, in contrast to the conventional introduction of quark matter through a first-order phase transition ({\it e.g.,} the MIT \cite{burgio2002maximum} and Nambu-Jones-Lasinio bag model approaches \cite{schertler1999neutron}).  The sound speed increase is, in fact, large enough to now allow the use of nucleon-nucleon potentials that couldn't otherwise satisfy the observational constraint that $M_{max}\gtrsim 2M_\odot$  \cite{ranea2015constant,montana2019constraining}. Fine-tuning of the quarkyonic or hadronic model parameters is not necessary in order to obtain viable models.

The model proposed by \citet{mclerran2019quarkyonic} is somewhat schematic, involving a single nucleon species and a chargeless 2-flavor (d and u) system of quarks without the consideration of chemical or beta equilibrium, which are the requirements that at every density, the energy is minimized with respect to particle and charge concentrations. The model also lacks protons and leptons (electrons and muons).  In fact, we will demonstrate that this chargeless 2-flavor approach cannot satisfy chemical or beta equilibrium. \citet{jeong2020dynamically} introduced a hard core potential for nucleons by means of an excluded volume, and reformulated {the quarkyonic model} so that it satisfies quark-nucleon chemical equilibrium. However, their model does not contain protons or leptons, and thus doesn't permit beta equilibrium between hadrons and leptons. { \citet{duarte2020excluded} have extended the excluded volume approach to take beta equilibrium into account.} 

In this paper, we propose a quarkyonic model {similar to that of \cite{mclerran2019quarkyonic} but} including protons and leptons with the essential elements to satisfy chemical and beta equilibrium that satisfies the experimental and observational constraints on neutron star structure.  
We describe the chargeless 2-flavor quark model in \S \ref{sec:cqm}, and then replace it in \S \ref{sec:mod} with a modified approach extended to include asymmetric nucleon matter and leptons.  In \S \ref{sec:eos} we point out the key features of the quarkyonic EOS, and in \S \ref{sec:constraint} we show how the available experimental and observational constraints can limit its parameter ranges.  \S \ref{sec:disc} examines the resulting semi-universal relations involving the dependence of neutron star structural quantities, such as the tidal deformability and binding energy, on the neutron star mass and radius, as well as implications for the direct Urca process. { We develop in Appendix \ref{app:simple} a simpler $ndu$ version of quarkyonic matter, without the complications of leptons or beta equilibrium, that is completely analytic, rendering it as a   particularly convenient and useful physically-motivated parameterized EOS for the interpretation of observational data.}

\section{Formulation of the Chargeless 2-Flavor Quarkyonic Model\label{sec:cqm}}
The chargeless 2-flavor quarkyonic model \cite{mclerran2019quarkyonic} assumes that strongly interacting quarks near the Fermi sea form interacting neutrons, while the remaining d and u quarks are non-interacting and fill the lowest momenta up to $k_{Fu}$ and $k_{Fd}$, respectively. The total baryon number density of quarkyonic matter is
\begin{eqnarray}
    &&n_B = n_n + \frac{N_c}{3} (n_u+n_d)\nonumber\\
    &=&{\frac{g_s}{2\pi^2}}\left[\int_{k_{0n}}^{k_{Fn}}k^2dk + {\frac{N_c}{3}}\left(\int_0^{k_{Fu}}k^2dk +\int_0^{k_{F{d}}}k^2dk\right)\right]\nonumber\\ &=&\frac{g_s}{6\pi^2}\left[ k_{Fn}^3-k_{0n}^3 +\frac{N_c}{3}\left(k_{Fu}^3+k_{Fd}^3\right)\right]
\label{eq:den}\end{eqnarray}
where $k_{Fn}$, $k_{Fu}$, and $k_{Fd}$ are the Fermi momenta of neutrons and u and d quarks, respectively.  Fermion spin degeneracy and quark color degeneracy are $g_s=2$ and $N_c=3$, respectively.  The neutrons are restricted to momentum states near the Fermi surface by the introduction of $k_{0n}$, the minimum allowed neutron momentum. It was arbitarily assumed that 
\begin{equation}
k_{0n}=N_ck_{Fd}=k_{Fn}\left[1-\left({\frac{\Lambda}{\hbar  k_{Fn}c}}\right)^3-{\frac{\kappa\Lambda}{N_c^2\hbar k_{Fn}c}}\right]
\label{eq:qrel}\end{equation}
so that $n_n$ and $n_d$ can both be written solely as functions of $k_{Fn}$.  To preserve charge neutrality, it was assumed 
\begin{equation}
   k_{Fd}=2^{1/3}k_{Fu}, 
\label{eq:qcons}\end{equation}
so that $n_u$ also can also be written solely in terms of $k_{Fn}$.
Note that the specification of a transition density $n_t$ determines the parameter $\kappa$ via
\begin{equation}
    \kappa=9\left[{\hbar k_{tn}c\over\Lambda}-\left({\Lambda\over\hbar k_{tn}c}\right)^2\right],\label{eq:kappa}
\end{equation}
where the corresponding neutron Fermi momentum at $n_t$ is $k_{tn}=(3\pi^2n_t)^{1/3}$.
 It is interesting to note that in this model, using ${ g_s}=2$ and $N_c=3$,
\begin{eqnarray}
    n_B&=&{1\over3\pi^2}\left(k_{Fn}^3{-{51\over2}}k_{Fd}^3\right),\nonumber\\
{k_{Fd}}&=&{k_{Fn}-k_{tn}\over3}\left[1+{\Lambda^3(k_{tn}+k_{Fn})\over(\hbar c)^3 k_{tn}^2 k_{Fn}^2}\right],
\end{eqnarray}
and both {$n_B$ and $k_{Fd}$} are explicit functions of $k_{Fn}$.

 The energy density of interacting neutrons is the sum of their relativistic kinetic and potential energy densities 
\begin{multline}
    \varepsilon_n(k_{Fn},k_{0n})=\frac{g_s}{2\pi^2} \int ^{k_{Fn}} _{k_{0n}}  k^2\sqrt{m_B^2c^4+(\hbar kc)^2}dk\\
        + n_n V(n_n),
    \label{eq:eden}   
    \end{multline}
    where $m_B$ is the baryon mass and $V(n_n)$ is the neutron potential energy, assumed to depend only on $n_n$.  It is assumed that the potential energy function remains the same both below and above the density $n_t$ where free quarks start to appear. It turns out that the particular choice of nucleon potential is not crucial to achieve the goals of achieving both a large maximum mass and reasonable neutron star radii, given allowable variations in the parameters $n_t$ and $\Lambda$.   

    The quark energy densities are given by the expressions for non-interacting fermions, 
    \begin{equation}
 \varepsilon_{d,u}=\frac{g_sN_c}{2\pi^2}\int_0^{k_{F(d,u)}}k^2\sqrt{m_{d,u}^2c^4+(\hbar kc)^2}\, dk.
   \label{eq:nif} \end{equation}
It is apparent that, with the use of Eqs. (\ref{eq:qrel}) and (\ref{eq:qcons}), the total energy density $\varepsilon=\varepsilon_n+\varepsilon_d+\varepsilon_u$ can also be written as a function of $k_{Fn}$ alone.
    Defining the baryon chemical potential by $d\varepsilon/dn_B=\mu_B$, the total pressure and sound speed are
 \begin{equation}
     p=n_B\mu_B-\varepsilon,\qquad
      \frac{c_s^2}{c^2}=\frac{1}{\mu_B}\frac{dp}{dn_B}.\end{equation}  
 
It is straightforward to demonstrate that this model cannot satisfy energy {optimization} with respect to the constituent compositions $n_n, n_d$ and $n_u$, all of which are functions of $k_{Fn}$ alone.  Minimizing the total energy density {at fixed $n_B$} with respect to the density of neutrons, d quarks or u quarks, {\it i.e.,} with respect to $k_{Fn}$, { becomes} tantamount to setting the baryon chemical potential $\mu_B$ equal to zero, which is contradictory.  Therefore,  neutrons and quarks are not in chemical equilibrium. {The two conditions involving $k_{0n}$ of Eq. (\ref{eq:k0}) over-constrain the system.} If  extended to include protons and leptons, {the energy of }this particular model could also not be { optimimized} with respect to the charge densities, which describes beta equilibrium.

\section{The Modified Quarkyonic Model\label{sec:mod}}

We seek a model that allows for energy minimization and will also include leptons as well as protons.  Leptons in beta equilibrium consist of both electrons and muons as long as $n_t\gtrsim n_s$, the nuclear saturation density 0.16 fm$^{-3}$, which is believed to be the case owing to the lack of experimental information indicating otherwise.   
The total energy density is
\begin{equation}
    \varepsilon = \varepsilon_B +\varepsilon_e+\varepsilon_\mu +\varepsilon_d +\varepsilon_u,
\end{equation}
 where $\varepsilon_B$ is the energy density of interacting neutrons and protons, while the energy densities of the leptons and quarks are given by Eq. (\ref{eq:nif}) for non-interacting fermions, 
using $g_s=2$, $N_c=3(1)$ for quarks (leptons), and the appropriate Fermi momenta and masses.
The lepton number densities are $n_e=k_{Fe}^3/(3\pi^2)$ and $n_\mu=k_\mu^3/(3\pi^2)$, and the total pressure is $p=n_B\mu-\varepsilon$ {where $\mu=d\varepsilon/dn_B$}.

We will describe the interactions among nucleons with a nucleon potential energy that depends on both neutron and proton densities $n_n$ and $n_p$, respectively.  We write this in terms of 6 sub-parameters $[a_0,b_0,a_1, b_1,\gamma,\gamma_1]$ {fit to} selected properties of uniform nucleonic matter.  We take the symmetry contribution to be adequately described by retention of just the lowest order quadratic term in the neutron excess.  We assume
\begin{eqnarray}
    V(n_n,n_p)&=&4x(1-x)\left(a_0u+b_0u^\gamma\right)\nonumber\\
&+&(1-2x)^2\left(a_1u+b_1u^{\gamma_1}\right),
    \label{eq:pot}
\end{eqnarray}
where $u=(n_n+n_p)/n_s$ and $x=n_p/(n_n+n_p)${; $V(n_n,n_p=0)=V(n_n)$}.
For symmetric nuclear matter {(SNM)}, the fitted quantities are the bulk binding energy $B\simeq16$ MeV, pressure ($p_B=0$), and incompressibility parameter $K_{1/2}\simeq220$ MeV at $n_s$.  For pure neutron matter {(PNM)}, the fitted quantities are its energy (relative to the baryon mass $m_B=939.5$ MeV) $E_0=S_V-B\simeq15$ MeV and pressure 1.6 MeV fm$^{-3}\lesssim p_0=Ln_s/3\lesssim4.0$ MeV fm$^{-3}$ evaluated at the same density $n_s$.  $S_v\simeq31$ MeV and 30 MeV$\lesssim L\lesssim70$ MeV are the usual nuclear symmetry energy parameters.  Since $L$ is by far the most uncertain of the fitted nuclear quantities, we choose it to be the only nucleonic free parameter.  As is well known, the parameter $L$ has a strong correlation with the intermediate-mass (e.g., $1.4M_\odot$) neutron star radius $R_{1.4}$.

The sub-parameters are determined using
\begin{eqnarray}
    \gamma&=&{K_{1/2}/9-T^{\prime\prime}_{1/2}\over T_{1/2}-T_{1/2}^\prime+B},\quad
    b_0={K_{1/2}/9-T^{\prime\prime}_{1/2}\over\gamma(\gamma-1)},\nonumber\\
    a_0&=&-B-T_{1/2}-b_0,\quad b_1={L/3+B-S_v+T_0-T_0^\prime\over\gamma_1-1},\nonumber\\
    &{\rm and}&\quad a_1=S_v-B-T_0-b_1,\label{eq:par}
\end{eqnarray}
where $T_{1/2}\simeq21.79$ MeV, $T^\prime_{1/2}${$=udT_{1/2}/du$}$\simeq14.34$ MeV and $T^{\prime\prime}_{1/2}${ $=u^2d^2T_{1/2}/du^2$}$\simeq-5.030$ MeV are the {SNM} kinetic energy and its first two logarithmic derivatives evaluated at $n_s$.  $T_0\simeq34.33$ MeV and $T^\prime_0${$=udT_0/du$}$\simeq22.41$ MeV are the PNM kinetic energy and its first logarithmic derivative evaluated at the same density.  $T_{1/2}$ and $T_0$ are taken relative to $m_B$.  The parameters $\gamma\simeq1.256, a_0\simeq-129.3$ MeV and $b_0=91.49$ MeV are determined by properties of {SNM}, but don't depend on the symmetry parameters $S_v$ or $L$.  $a_1$ and $b_1$ are sensitive to $S_v$, $L$ and $\gamma_1$, on the other hand.  We obtain $a_1\simeq -L/2-14.70$ MeV and $b_1\simeq L/2-4.63$ MeV for the choices $S_V=31$ MeV and $\gamma_1=5/3$. 

\begin{figure}[ht]
\centering\includegraphics[width=1.\linewidth,angle=0]{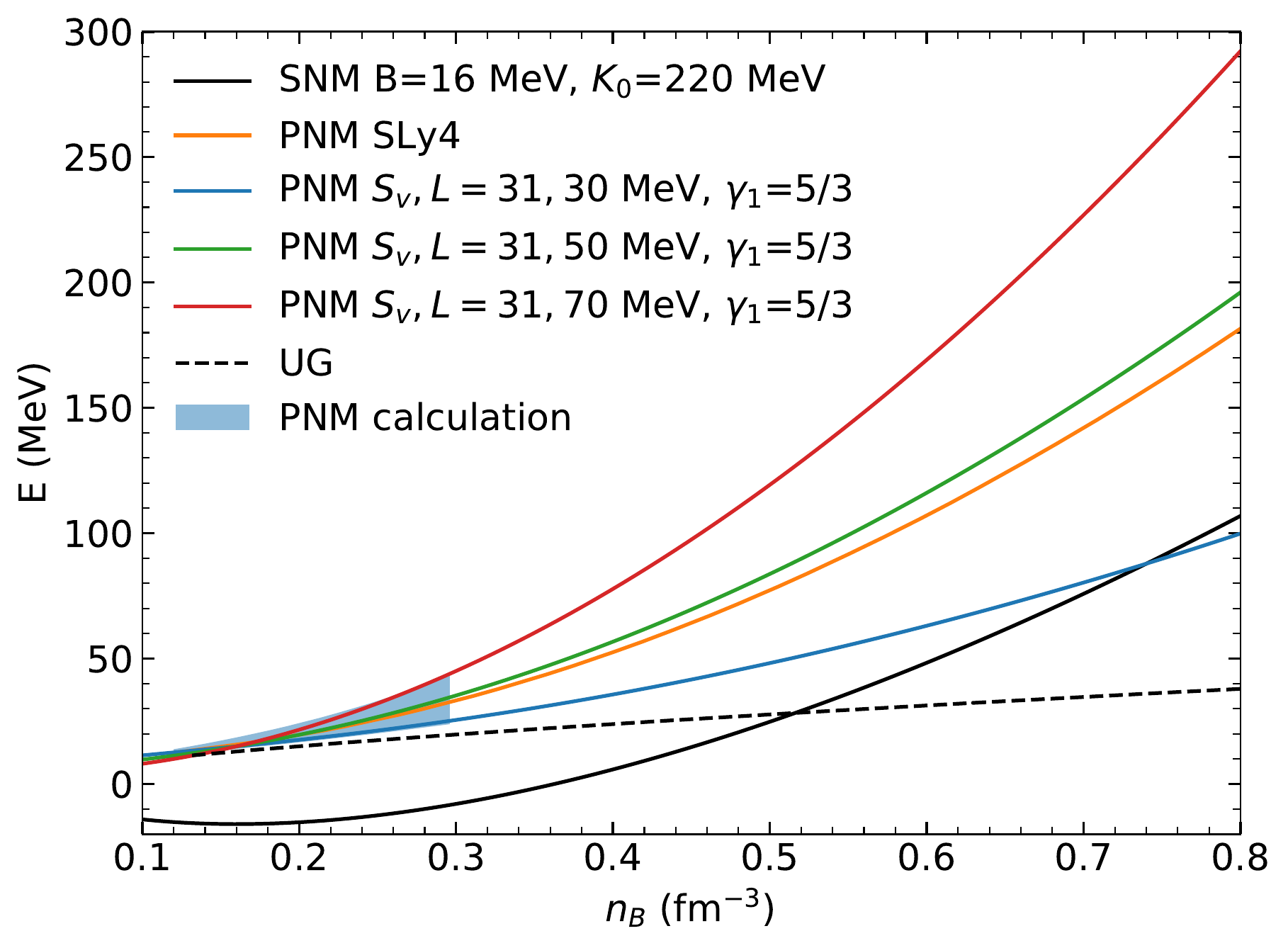}
\caption{The energy per baryon {$E$} of symmetric nuclear matter (SNM - black), and the energy per baryon of pure neutron matter (PNM) for the standard parameter set (see text), for three values of $L$ (30 MeV - yellow, 50 MeV - green, 70 MeV - red).  The blue curve shows the crust equation of state, SLy4, continued to high densities. The dashed curve shows the {conjectured} unitary gas bound \cite{tews2017symmetry}. The shaded region indicates PNM ranges from theoretical models \cite{gandolfi2012maximum}. \label{fig:EPB}}
\end{figure}

 The total energy density $\varepsilon_B$ of interacting nucleons is
\begin{multline}
    \varepsilon_B=\sum_{i=n,p} \frac{g_{s}}{2\pi^2} \int ^{k_{Fi}} _{k_{0i}} k^2\sqrt{m_B^2c^4+(\hbar kc)^2}\, dk \\ +(n_n+n_p) V(n_n,n_p).
\label{eq:enuc}\end{multline}
Fig. \ref{fig:EPB} shows the specific nucleon energy, $\varepsilon_B/n_B-m_B$, as a function of density for { SNM and PNM}, assuming the standard values $n_s=0.16$ fm$^{-3}$, $B=16$ MeV, $K_{1/2}=220$ MeV, $S_v=31$ MeV, $L=30, 50$ and $70$ MeV, and $\gamma_1=5/3$.   The choice of $\gamma_1=5/3$ gives an excellent representation of recent chiral Lagrangian {PNM} calculations \cite{gandolfi2012maximum} for baryon densities $0.5n_s<n_B<1.8n_s$ (the shaded region in Fig. \ref{fig:EPB}).  $L=30$ MeV and $L=70$ MeV are seen to bracket the theoretical PNM results. The choice $L=30$ MeV is also consistent with the theoretical minimum for the PNM energy predicted by the {\it unitary gas constraint} \cite{tews2017symmetry}.  The neutron matter energy is larger than the symmetric matter energy for all densities except at densities $n_B\gtrsim0.7$ fm$^{-3}$ in the case of a very soft symmetry energy, i.e., $L=30$ MeV. A more complex SNM parameterization, for example one that can also fit its skewness at $n_s$, could prevent this.  This affects our results only marginally, however. 

  In the neutron star crust, at densities below $n_{cc}\simeq0.07$ fm$^{-3}$, we assume the SLy4 equation of state \cite{chabanat1998skyrme}, which nearly matches the PNM energies at $n_{cc}$ for all $L$ values considered.  For comparison, the SLy4 EOS continued to high densities is shown in Fig. \ref{fig:EPB}.   Our results are not sensitive to the precise choice of the crust EOS.

With the addition of d and u quarks, the general conditions of baryon and charge conservation are
\begin{eqnarray}
    n_B&=&n_n+n_p+{n_d+n_u\over3},\\ n_BY_L&=&n_e+n_\mu=n_p+{2n_u{-n_d}\over3},
\label{eq:cons}\end{eqnarray}
where $Y_L$ is the net lepton fraction.

The modified quarkyonic model retains the paradigm that nucleons are restricted to momentum shells near the Fermi surface for $n_B>n_t$.  However, the imposition of the constraint $k_{0n}=3k_d$ is an additional relation between $k_{Fn}$ and $k_{Fd}$ that is incompatible with energy minimization with respect to particle compositions.  We therefore abandon and replace it with the condition of chemical equilibrium among the nucleons and quarks.  {\it This is the key difference between our model and that of \citet{mclerran2019quarkyonic}}.

Since protons are now considered, we introduce the minimum momentum for protons in quarkyonic matter, $k_{0p}${, such that $n_{n,p}=g_s(k_{F(n,p)}^3-k_{0(n,p)}^3)/(6\pi^2)$}.  Both $k_{0n}$ and $k_{0p}$ are assumed to be functions only of their corresponding Fermi momenta by the relations
\begin{equation}
    \hspace*{-.3cm}k_{0(n,p)}=k_{F(n,p)}\left[1-\left({\Lambda\over\hbar k_{F(n,p)}c}\right)^2-{\kappa_{n,p}\Lambda\over9\hbar k_{F(n,p)}c}\right],
\label{eq:mod1}\end{equation}
which involve the parameters $\kappa_n$ and $\kappa_p$.  Note that these functions are modified from Eq. (\ref{eq:qrel}) by a change of exponent which forces $k_{0(n,p)}$ to change more slowly when quarkyonic matter appears, making for more stable solutions in this regime.  This assumption does not affect the ability of this model to give rise to a significant increase in the sound speed near the transition density.  

It is also assumed that momenta restrictions on neutrons and protons begin above a common transition density $n_t$, which determines $\kappa_{n,p}$:
\begin{equation}
    \kappa_{n,p}=9\left[{\hbar k_{t(n,p)}c\over\Lambda}-{\Lambda\over\hbar k_{t(n,p)}c}\right].\label{eq:mod2}
\end{equation}
The transition Fermi momenta $k_{t(n,p)}$ are obtained from beta equilibrium of the uniform $n,p,e,\mu$ system at the density $n_t$. Eliminating the $\kappa$ parameters, Eq. (\ref{eq:mod1}) is reformulated as
\begin{equation}
    k_{0(n,p)}=(k_{F(n,p)}-k_{t(n,p)})\left[1+{\Lambda^2\over(\hbar c)^2k_{F(n,p)}k_{t(n,p)}}\right].
\label{eq:k0}\end{equation}
For sufficiently low values of $L$ MeV, it is possible that {the symmetry energy can become negative  at high densities in hadronic matter, in which case} $dk_{Fp}/dn_B<0$.  Depending on the parameters, this can result in a situation in which $k_{Fp}<k_{tp}$ in quarkyonic matter above transition densities between $0.3 (0.6)$ fm$^{-3}$ (for $L=30$ MeV and $\Lambda=500 (1700)$ MeV) and $0.6 (1.2)$ fm$^{-3}$ (for $L=40$ MeV and $\Lambda=500 (1700)$ MeV).  { And if} $k_{Fp}$ were to fall to zero before the quarkyonic sector is reached at $n_B=n_t$, then { the lowest energy state would be PNM, dictating} that $k_{Fp}=0$ at all higher densities including in the quarkyonic sector.  If $k_{Fp}$ is positive when quarks appear, when $k_{Fp}<k_{tp}$ we instead require $k_{0p}=0$ in the quarkyonic sector because otherwise Eq. (\ref{eq:k0}) becomes ill-defined{, and this simply means the proton Fermi shell is not 'saturated'}.  We could have chosen a slightly more complex SNM energy, such that $E_{SNM}<E_{PNM}$ at all densities, that would avoid these situations yet would have little effect on the results of this paper {because the proton fraction is small.}  { See the Appendix for a simpler formulation that avoids this issue.}

Strong interaction equilibrium dictates total energy minimization with respect to particle concentrations at fixed density and lepton fraction and is valid under nearly all circumstances in astrophysical simulations.  This condition is equivalent to chemical equilibrium among the nucleons and quarks, and leads to 
\begin{equation}
   \mu_d={2\over3}\mu_n-{1\over3}\mu_p,\qquad\mu_u={2\over3}\mu_p-{1\over3}\mu_n.
\label{eq:qeq}\end{equation}
These relations replace expressions of the form $k_{0(n,p)}\propto k_{d,u}$ that are integral to the model of \cite{mclerran2019quarkyonic}.  The nucleon chemical potentials are (see also {\cite{han2019treating}})
\begin{eqnarray}
&&\hspace*{.5cm}\mu_{n,p}={\partial\varepsilon_B\over\partial n_{n,p}}= \left(1-K_{n,p}\right)^{-1}\times\nonumber\\
&\times&\left(\sqrt{m_B^2c^4+(\hbar ck_{F(n,p)})^2}-K_{n,p}\sqrt{m_B^2c^4+(\hbar ck_{0(n,p)})^2}\right)\nonumber\\
&&\hspace*{1cm}+{\partial [(n_n+n_p)V(n_n,n_p)]\over\partial n_{n,p}}, \label{eq:chempo_neutron}
\end{eqnarray}
where
\begin{eqnarray}
    K_{n,p}&=&\left({k_{0(n,p)}\over k_{F(n,p)}}\right)^2{dk_{0(n,p)}\over dk_{F(n,p)}}\nonumber\\
    &=&\left({k_{0(n,p)}\over k_{F(n,p)}}\right)^2\left[1+\left({\Lambda\over\hbar ck_{F(n,p)}}\right)^2\right].
\label{eq:K}\end{eqnarray}

\begin{figure}[ht]
\includegraphics[width=\linewidth,angle=0]{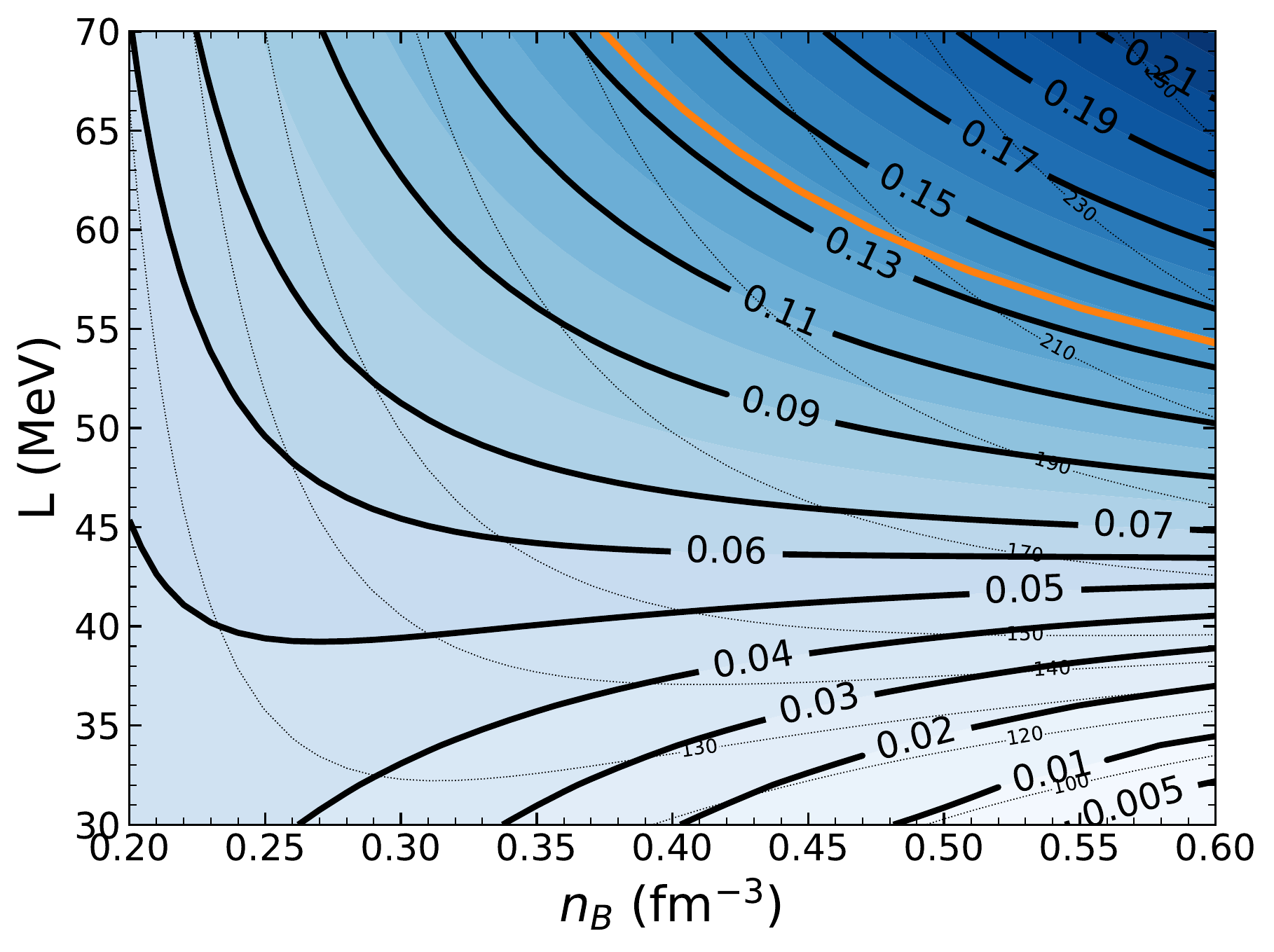}
\caption{The lepton fraction $Y_L$ of pure hadronic matter in beta-equilibrium as functions of $n_B$ and $L$.   Overlain are contours of $\mu_{tn}-\mu_{tp}=m_d-m_u$.  Also shown is the threshold lepton fraction for operation of the nucleon direct Urca neutrino cooling process. }
\label{fig:hbeta}\end{figure}

The condition of beta, or weak-interaction, equilibrium  is equivalent to the additional minimization of the total energy density with respect to the lepton concentrations at fixed baryon density, which applies if weak-interaction timescales are short compared to dynamical timescales.   Under conditions of charge neutrality, this gives
\begin{equation}
    \mu_e=\mu_\mu=\mu_n-\mu_p=\mu_d-\mu_u.
\label{eq:hbeta}\end{equation} 
One also finds that the effective chemical potential $\mu=d\varepsilon/dn_B=\mu_n$ under conditions of chemical and beta equilibrium.  In matter composed of hadrons alone, the beta equilibrium depends on the nucleon potential; values of the lepton number, $Y_L$, are shown in Fig. \ref{fig:hbeta} as a function of density and the single nucleon parameter $L$.

The requirement that both flavors of quarks appear at the same density $n_t$ means that the quark masses, like $\kappa_{n,p}$, are not free model parameters.  Their values are determined by the ambient beta-equilibrium conditions at $n_t$ and therefore depend on the nucleon potential.  They are
found from Eq. (\ref{eq:qeq}) using $k_{td}=k_{tu}=0$:
\begin{equation}
    m_d={2\over3}\mu_{tn}-{1\over3}\mu_{tp},\qquad m_u={2\over3}\mu_{tp}-{1\over3}\mu_{tn},
\end{equation}
where $\mu_{t(n,p)}$ are the beta-equilibrium values of the chemical potentials in $np\mu e$ matter at $n_t$.  Both masses are of  order $m_B/3$, as expected for constituent quark masses, but $m_d-m_u=\mu_{tn}-\mu_{tp}$, which depends on $n_t$ and $V$ ({\it i.e.,} $L$), and is in the range 80-250 MeV (Fig. \ref{fig:hbeta}).

\begin{figure}[ht]
\includegraphics[width=\linewidth,angle=0]{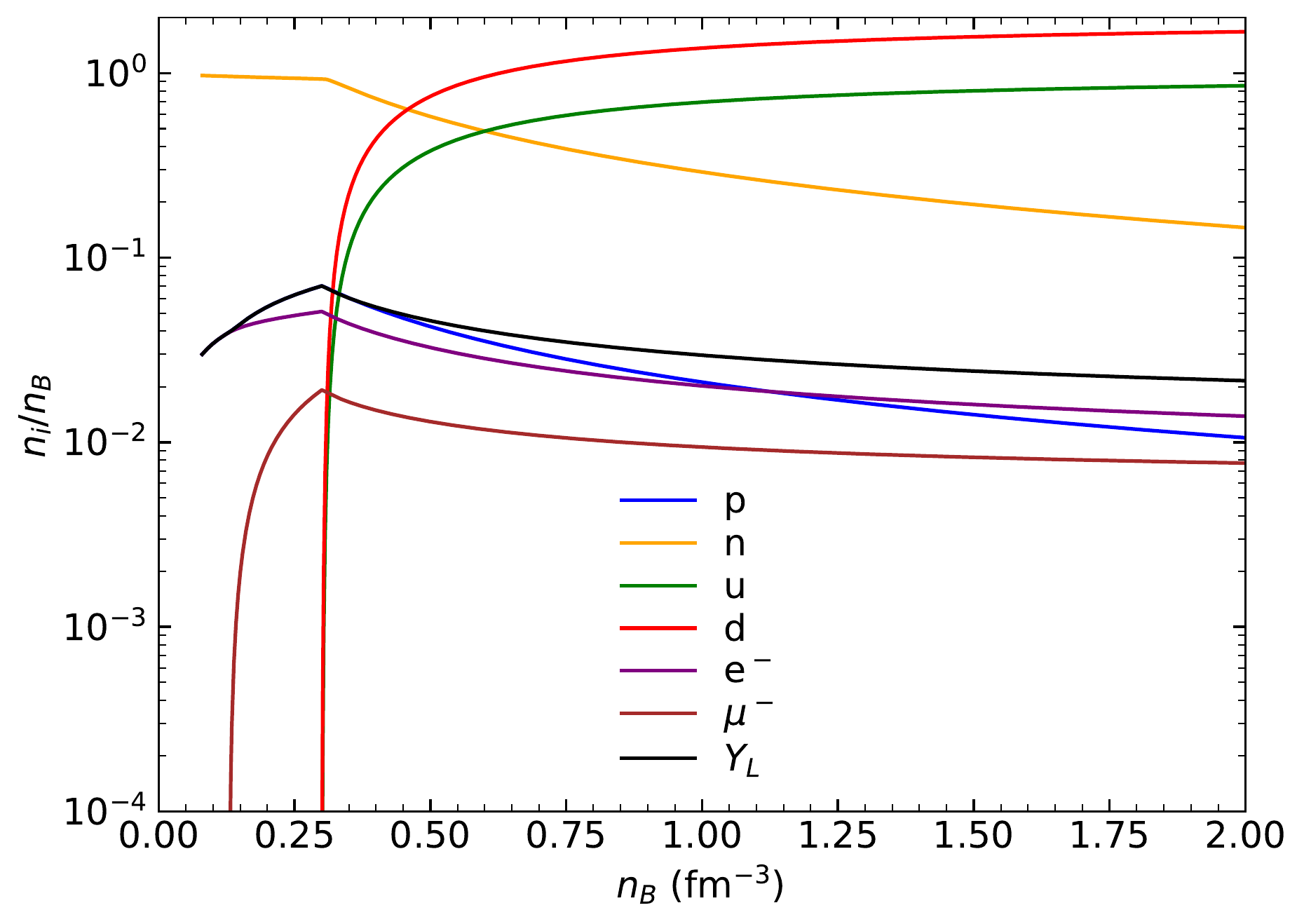}
\caption{Particle fractions in beta equilibrium with L=50 MeV, $\Lambda=1400$ MeV, and $n_t=0.3$ fm$^{-3}$.  The black line show the lepton fraction $Y_L$.}
\label{fig:partfrac}\end{figure}

 \begin{figure}[ht]
\includegraphics[width=\linewidth,angle=0]{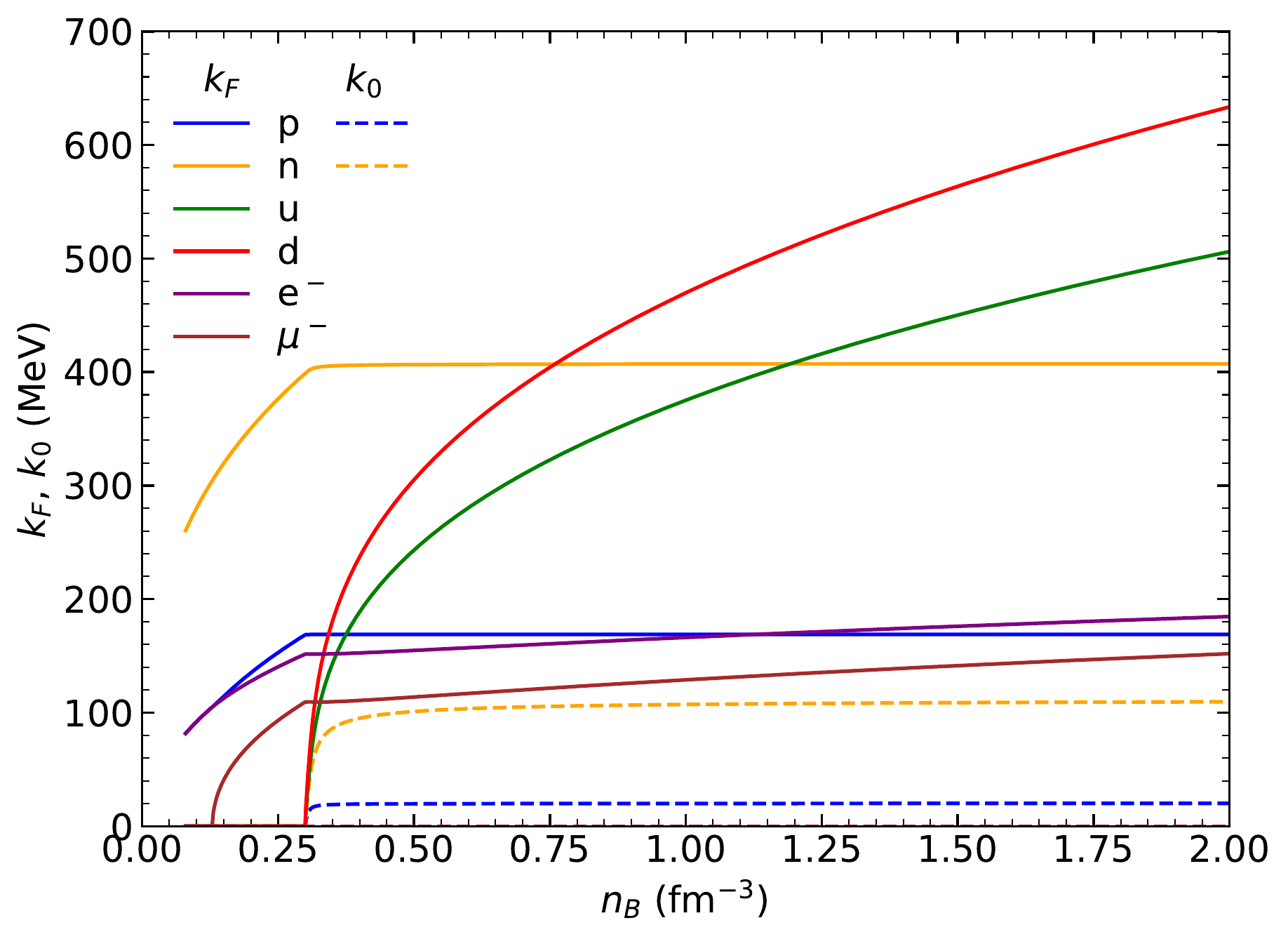}
\caption{Particle wavenumbers for quarkyonic matter in beta equilibrium with $L=50$ MeV, $\Lambda=1400$ MeV, and $n_t=0.3$ fm$^{-3}$.  Additionally are shown the minimum momenta $k_{0(n,p)}$ for nucleons in the quarkyonic sector.}
\vspace*{-.5cm}\label{fig:wave}\end{figure}
The appearance of quarks for $n>n_t$ drastically alters the composition.  Fig. \ref{fig:partfrac} shows the particle fractions in beta-equilibrium matter for a standard case with $\Lambda=1400$ MeV, $n_t=0.3$ fm$^{-3}$ and $L=50$ MeV. 
For this case $\kappa_p=-74.54$,
 $\kappa_n=-29.00$, $m_u=241.07$ MeV, and $m_d=391.28$ MeV.
  The corresponding Fermi wavenumbers and minimum nucleon wavenumbers $k_{0(n,p)}$ in the quarkyonic sector are displayed in Fig. \ref{fig:wave}.

For densities slightly in excess of $n_t$, the total nucleon density initially varies linearly with $n_B-n_t\equiv\delta$, which follows from baryon conservation since the quark densities can be ignored (as we show below).  The proton fraction $n_p/(n_n+n_p)$ decreases linearly (but slowly) with increasing $\delta$ because $k_{0(n,p)}$ and $k_{F(n,p)}-k_{t(n,p)}$ also vary linearly with $\delta$, and the neutron momenta $k_{0,n}$ and $k_{F,n}$ vary more rapidly than their proton counterpart.  In comparison, $n_d$ and $n_u$ are initially small. It is found that, irrespective of the value of $\Lambda$, the quark momenta scale as $\sqrt{\delta}$ for small $\delta$, so that $n_{d,u}\propto\delta^{3/2}$.  

However, both $k_{Fn}$ and $k_{Fp}$ saturate in the quarkyonic sector with values near their values $k_{t(n,p)}$ at the transition density {as long as $\Lambda>\hbar ck_{(n,p)}$, which is shown in Sec. \ref{sec:constraint} to be necessary to satisfy the requirement that $M_{max}\gtrsim2M_\odot$}. This can be understood as a consequence of the fact that there is a maximum Fermi wavenumber $k_{m(n,p)}$ in the quarkyonic sector.  This is determined by
\begin{equation}
    {dn_{n,p}\over dk_{F(n,p)}} ={k_{F(n,p)}^2\over\pi^2}\left(1-K_{n,p}\right)=0,
\end{equation}
or simply $K_{n,p}=1$.  Using Eq. (\ref{eq:K}), one finds
\begin{equation}
    k_{m(n,p)}=k_{0(n,p)}\sqrt{1+\left({\Lambda\over\hbar ck_{m(n,p)}}\right)^2}.
\label{eq:kmax}\end{equation}
With Eq. (\ref{eq:k0}), and since $k_{F(n,p)}-k_{t(n,p)}<<k_{t(n,p)}$,
\begin{eqnarray}
k_{m(n,p)}\!\!\!&-&\!\!\!k_{t(n,p)}=k_{m(n,p)}\left[1+\left({\Lambda\over\hbar ck_{m(n,p)}}\right)^2\right]^{-1/2}\nonumber\\
&\times&\left[1+{\Lambda^2\over(\hbar c)^2k_{m(n,p)}k_{t(n,p)}}\right]^{-1}\\
\simeq\,\,&k_{t(n,p)}&\left({\hbar ck_{t(n,p)}\over\Lambda}\right)^3\left[1-{3\over2}\left({\hbar ck_{t(n,p)}\over\Lambda}\right)^2+\cdots\right],\nonumber\label{eq:kmax1}
\end{eqnarray}
where we kept up to the quadratic order terms of an expansion in $\hbar ck_{t(n,p)}/\Lambda$, which is generally much smaller than unity, in the last expression. Note that these relations depend only on $k_{t(n,p)}$ and $\Lambda$ and are valid for any nucleon potential ({\it i.e.,} $L$).
Obviously, $k_{0n,p}\sim k_{t(n,p)}(\hbar ck_{t(n,p)}/\Lambda)$ and $n_{n,p}\sim k_{t(n,p)}^3/(3\pi^2)$ also saturate.   Since $k_{tp}<k_{tn}$, the proton wavenumbers and density approach their asymptotic values before their neutron counterparts.  Therefore, the nucleon particle fractions must monotonically fall with increasing $n_B$ in the quarkyonic sector.    Because the nucleon fractions become small at high densities, the sound speed tends to $\sqrt{1/3}c$, the value implied by quark asymptotic freedom together with the relativistic behavior of the leptons.

Another property of the quarkyonic system in beta equilibrium observed in Fig. \ref{fig:wave} is that the quark Fermi momenta are nearly proportional to each other, with $k_{Fd}\simeq2^{1/3}k_{Fu}$ at all densities. This indicates that the quark and hadronic sectors are separately approximately charge neutral, despite the fact that global charge neutrality was imposed. This is reminiscent of the explicit assumption concerning quark momenta in Ref. \cite{mclerran2019quarkyonic}.  One also may note that $k_{Fd}$ increases more rapidly than $k_{0n,p}$, contrary to the assumption of  \citet{mclerran2019quarkyonic}, who assumed their ratio  to be fixed at $k_{0n}/k_{Fd}=N_c$.

\begin{figure}[ht]
\includegraphics[width=\linewidth,angle=0]{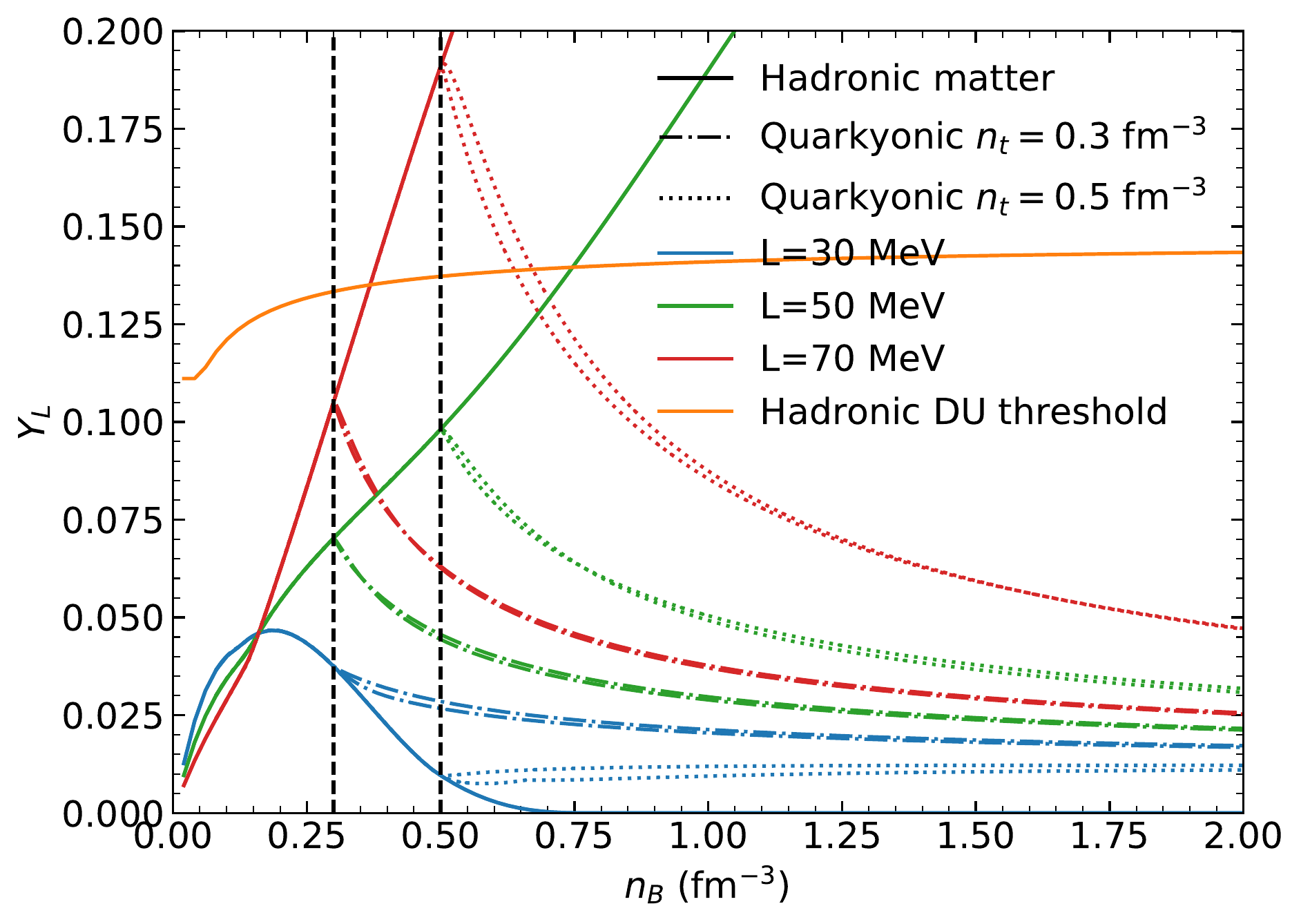}
\caption{Beta equilibrium lepton fraction as functions of model parameters $L=[30,50,70]$ MeV and  $n_t=[0.3,0.5]$ fm$^{-3}$.  The lower and upper members of each pair of curves have $\Lambda=800$ MeV and  1700 MeV, respectively). }
\label{fig:lhy}
\end{figure}
A comparison of the lepton fractions in beta equilibrium with three different nucleon potentials and two sets of standard quarkyonic parameters is shown in Fig \ref{fig:lhy}. The onset of quarks usually results in abrupt decreases in $Y_L$, with exceptions for large $n_t$ and small $L$.  It is apparent that the lepton fractions in the quarkyonic sector are insensitive to $\Lambda$.

 \section{The Equation of State\label{sec:eos}}
 
\begin{figure}[ht]
\includegraphics[width=\linewidth]{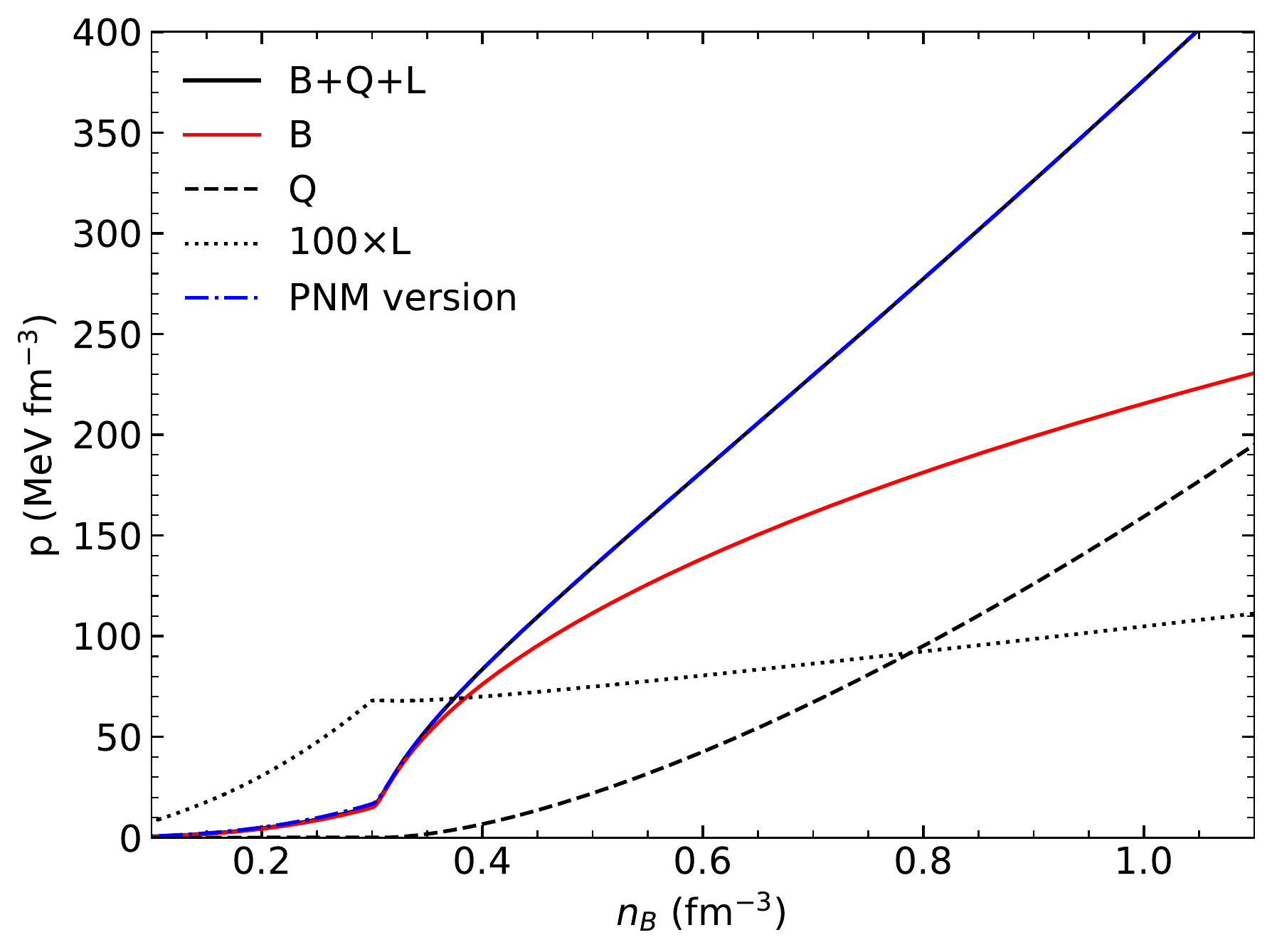}
\includegraphics[width=\linewidth]{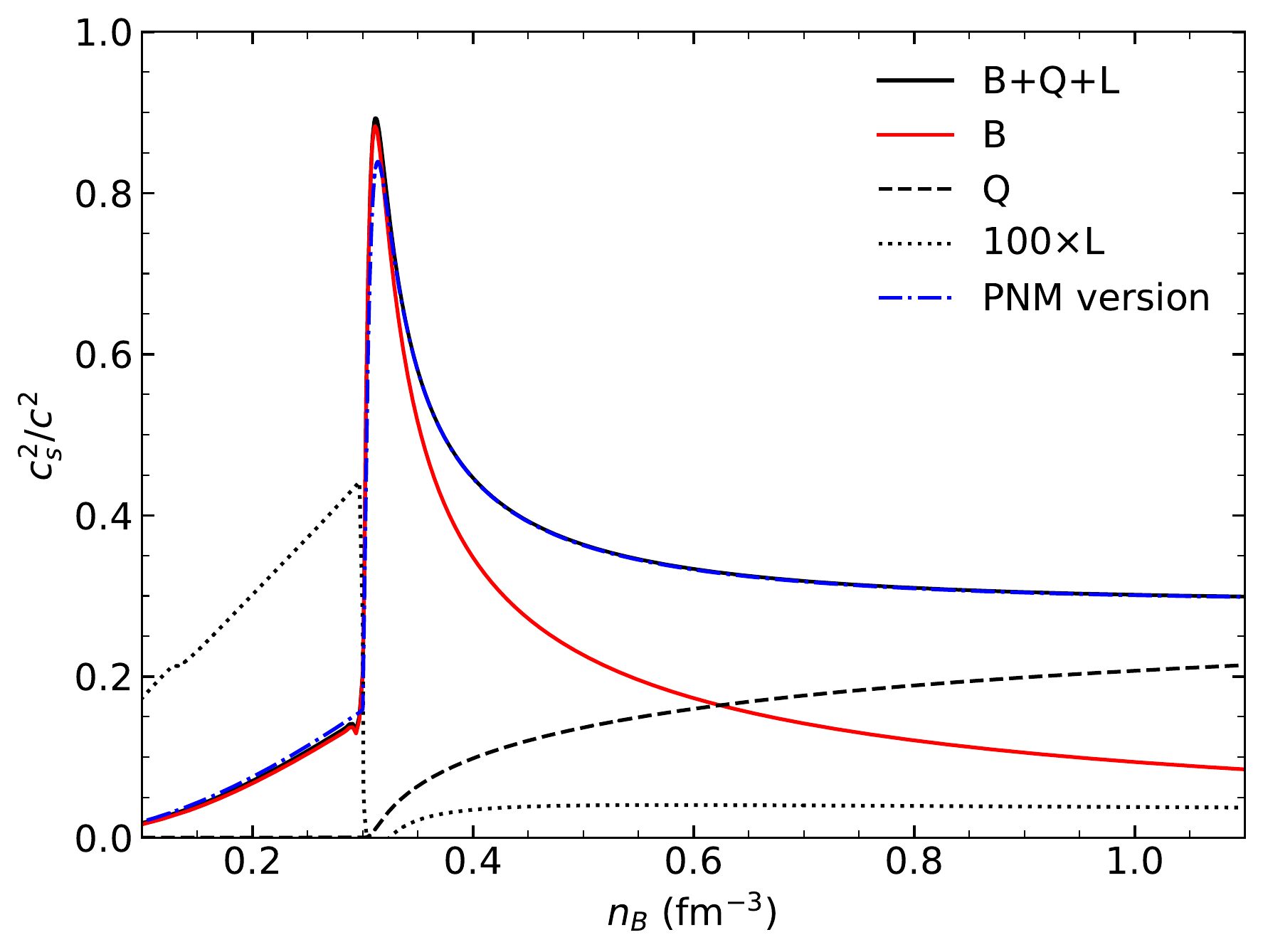}
\caption{The pressure $p$ (top panel) and sound speed $c_s$ (bottom panel) for the case $L=50$ MeV, $n_t=0.3$ fm$^{-3}$, and $\Lambda=1400$ MeV.  The totals as well as the individual contributions from baryons (B), leptons(L) and quarks (Q) are shown.  The lepton contributions are multiplied by 100 for clarity. }
\label{fig:tdet}\end{figure}
In contrast to conventional models of quark matter, for which the pressure initially remains constant or increases slowly with increasing density beyond $n_t$, the introduction of quarks in quarkyonic matter has a dramatic effect.  Details are shown for a particular parameter set ($L=50$ MeV, $n_t=0.3$ fm$^{-3}$, and $\Lambda=1400$ MeV) in Fig. \ref{fig:tdet}. 
The hadronic pressure is $p_B=\mu_nn_n+\mu_pn_p-\varepsilon_B$,
the leptonic pressure is $p_L=\mu_en_e+\mu_\mu n_\mu-\varepsilon_e-\varepsilon_\mu$ and the quark pressure is $p_Q=\mu_dn_d+\mu_un_u-\varepsilon_d-\varepsilon_u$. Breaking down the pressure $p=p_B+p_L+p_Q$ and its derivative in the form of the sound speed $c_s^2/c^2=\mu^{-1}d(p_B+p_L+p_Q)/dn_B$ into separate contributions from nucleons (B), leptons (L) and quarks (Q) shows that the rapid increase in $c_s$ is due to nucleons, through the restriction of their momenta.  As the quarks become more abundant, $c_s$ rapidly decreases due to their relativistic nature.
Leptons are relatively inconsequential.

\begin{figure*}
\includegraphics[width=0.495\linewidth,angle=0]{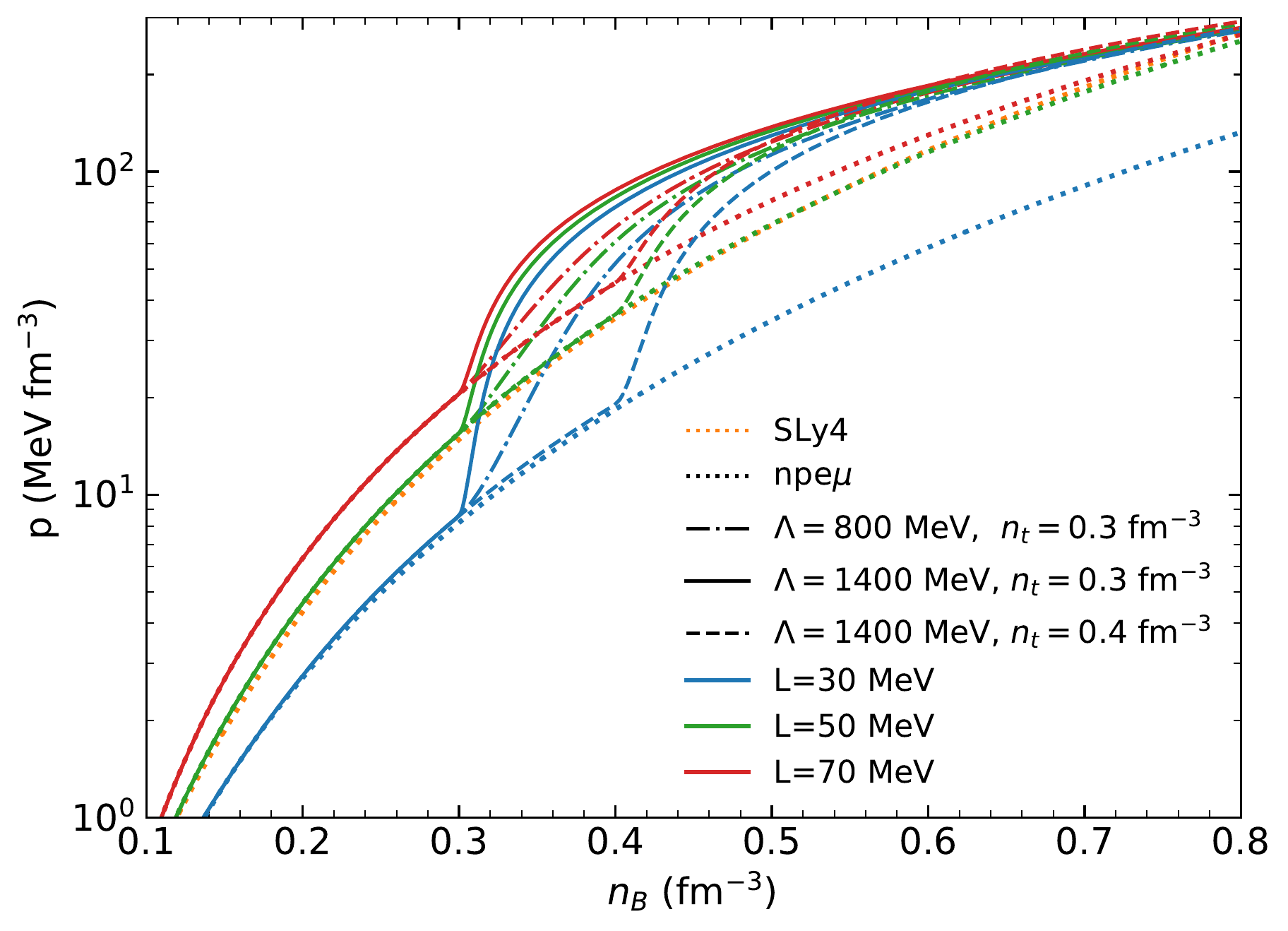}
\includegraphics[width=0.495\linewidth,angle=0]{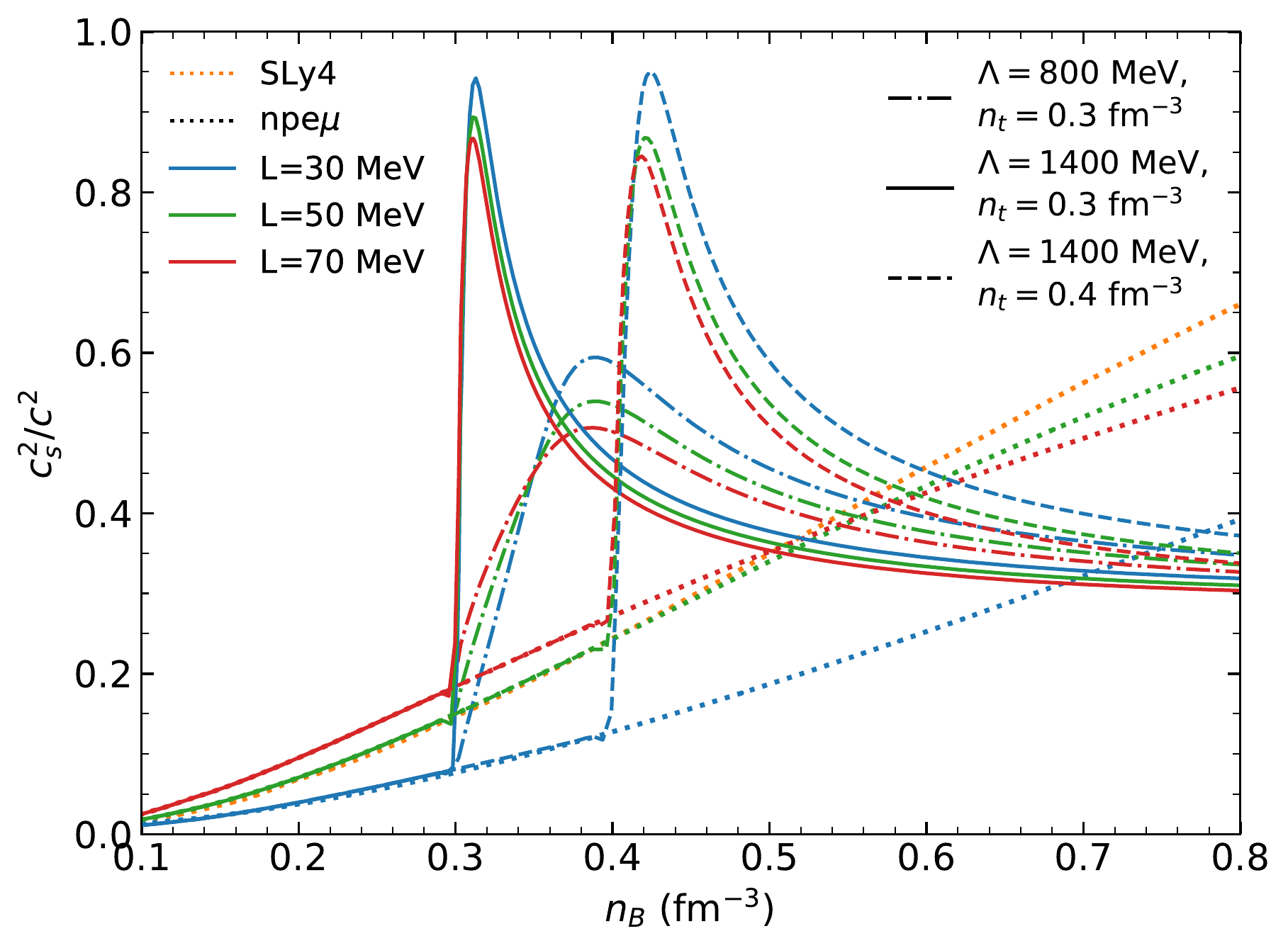}
\caption{The EOS for various combinations of model parameters, with the pressure $p$ displayed in the left-hand panel and sound speed $c_s$ in the right-hand panel.  Dotted lines show the SLy4 EOS continued to high densities as well as the three versions of the hadronic EOS with different $L$ values. }
\label{fig:eos}\end{figure*}
The EOS for several combinations of parameters are illustrated in Fig. \ref{fig:eos}.  The maximum sound speed is observed to be primarily a function of $\Lambda$.  The pressure of the quarkyonic EOS increases at all densities as $L$ is increased, but always approaches the asymptotic limit $p\sim\varepsilon/3$ for large $n_t$.

\begin{figure}
\centering\includegraphics[width= \linewidth,angle=0]{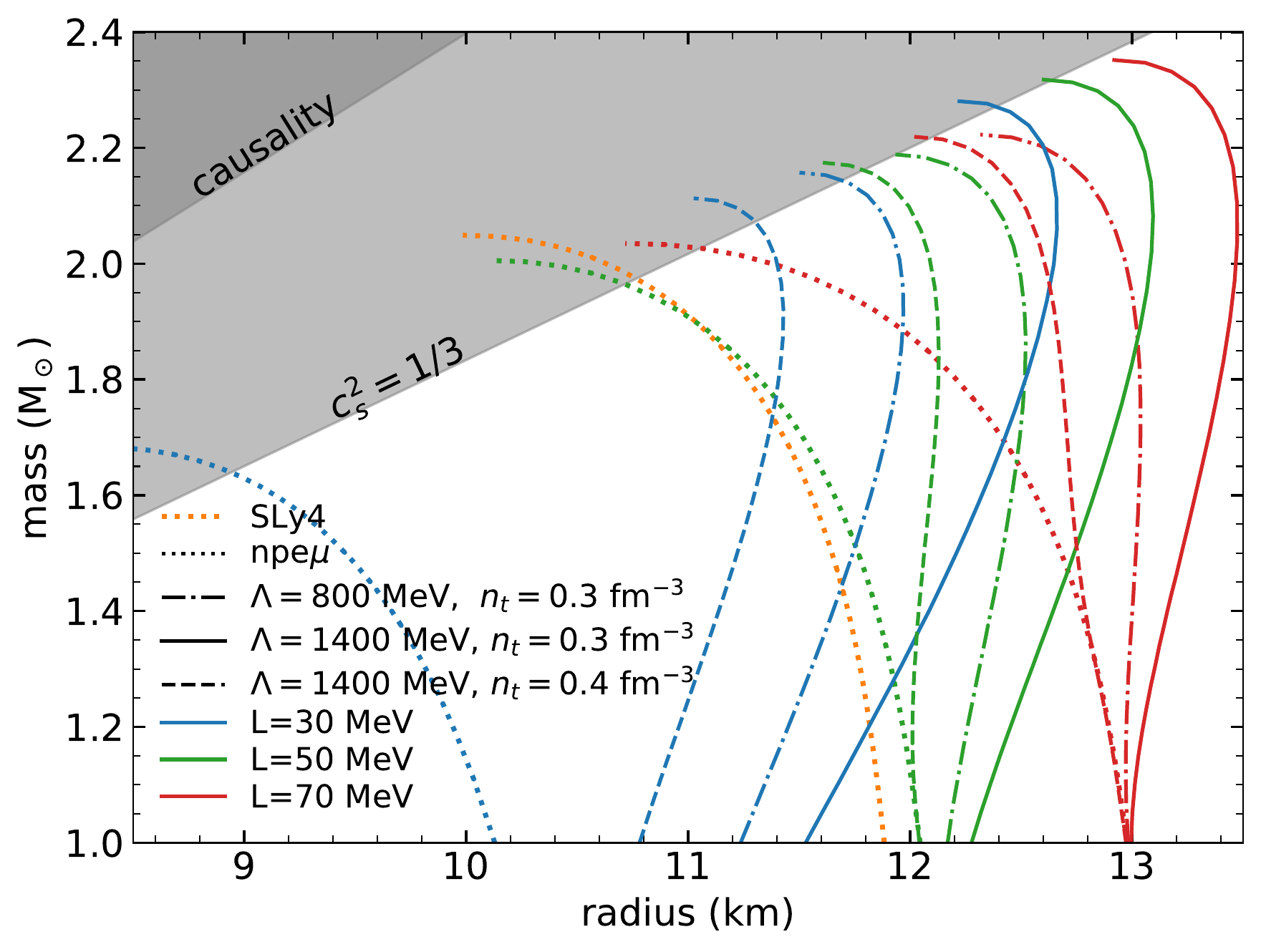}
\caption{Mass-radius curves for the same EOSs displayed in Fig. \ref{fig:eos}.  The dark shaded region is prohibited by causality, and the light shaded region would be prohibited if the sound speed was limited by $c_s^2\le c^2/3$. }
\label{fig:mr}\end{figure}
Mass-radius curves for the 13 EOSs displayed in Fig. \ref{fig:eos} are illustrated in Fig. \ref{fig:mr}.  Quarkyonic stars have larger radii for a given mass, and also larger maximum masses, than for the underlying hadronic EOS with the same value of $L$.  Generally, the radii for intermediate-mass stars increase with increasing $L$, but they also increase with increasing values of $\Lambda$ and decreasing values of $n_t$.  The fact that the pressure rapidly increases immediately beyond $n_t$ leads to the increases in radii of stars larger than $1M_\odot$ if $n_t\lesssim4n_s$. The radii of $1.4M_\odot$ stars is most sensitive to the pressure at approximately $2n_s$ \cite{Zhao2019}, but are still influenced by somewhat larger values of $n_t$.   

Note that the hadronic EOS predicted for $L=30$ MeV cannot achieve a maximum mass larger than about $1.7M_\odot$ (Fig. \ref{fig:mr}), yet quarkyonic models with $L=30$ MeV have no difficulty reaching masses in excess of $2.0M_\odot$ if $n_t\lesssim0.5$ fm$^{-3}$ (see \S \ref{sec:constraint}).  It is of note that a wide range of quarkyonic models can satisfy the GW170817 constraint suggesting $R_{1.4}\lesssim13.5$ km even with the largest value of $L$= 70 MeV that is able to satisfy theoretical neutron matter constraints in the vicinity of $n_s$.  
\section{Parameter Ranges Constrained by Causality and Neutron Star Observations\label{sec:constraint}}

\begin{figure}
\centering\includegraphics[width=\linewidth,angle=0]{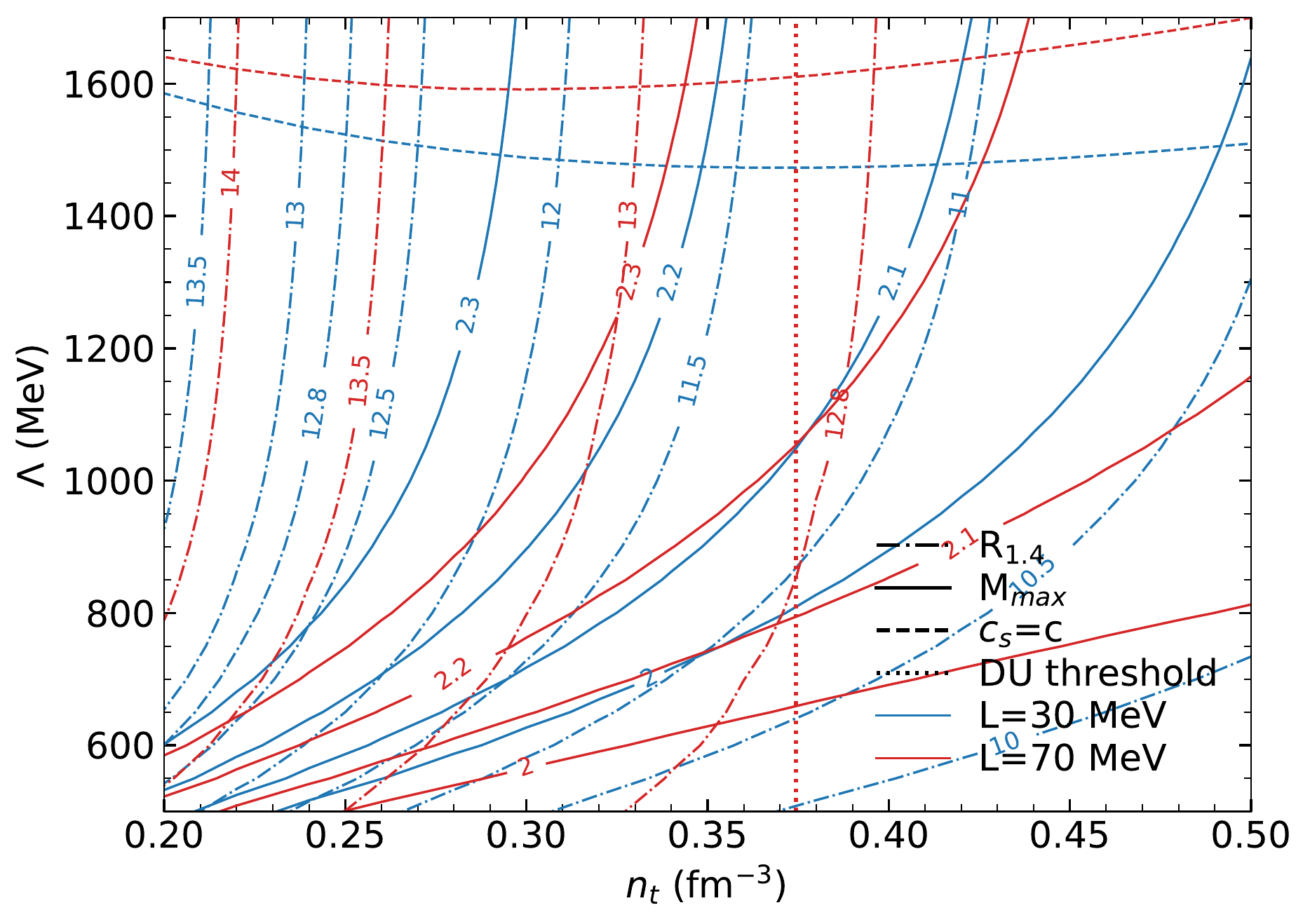}
\caption{Plausible parameter ranges.  Solid (dash-dotted) curves show maximum mass $M_{max}/M_\odot$ ($R_{1.4}$/km) contours for quarkyonic stars in beta equilibrium as functions of $\Lambda$ and $n_t$ for two extreme values of $L$, 30 MeV (blue) and 70 MeV (red).  The dashed contours indicate causality bounds. The dotted verical line demarks the DU threshold density for $L=70$ MeV. }
\label{fig:p-space}\end{figure}

Mass, radius and causality constraints on the model parameters are shown in Fig. \ref{fig:p-space}.  Radius constraints were discussed in \S \ref{sec:eos}.  The largest well-measured neutron star mass is PSR J0740+6620, for which $M=2.14^{+0.10}_{-0.09} M_\odot$ \citep{cromartie2020relativistic}. There are other measured masses which are smaller but have less uncertainty, such as  $M=2.01\pm0.04M_\odot$ for PSR J0438+0432 \citep{antoniadis2013massive}, and larger but with more uncertainty, such as $2.27^{+0.17}_{-0.15} M_\odot$ for PSR J2215-5135 \citep{linares2018peering}.  These collectively form a lower limit to the neutron star maximum mass $M_{max}$.  

A potential upper limit $M_{max}\lesssim2.3M_\odot$ was afforded by multi-messenger observations of the binary neutron star merger GW170817.  This provided evidence that the coalesced remnant initially formed a hypermassive neutron star which was partially supported by differential rotation.   The support from differential rotation briefly (perhaps a few tenths of a second) prevented the coalesced remnant from immediately collapsing into a black hole, which subsequently occurred if the remnant had a mass $M_{rem}$ greater than the maximum mass that could have been supported by uniform rotation at the Keplerian (mass-shedding) limit, about $M_{max,u}\simeq1.17M_{max}$.  An immediate collapse would have short-circuited not only the observed gamma-ray burst, which occurred 1.7 seconds after the gravitational wave event, but also the extensive mass ejection revealed by the appearance of the subsequent kilonova.  The maximum mass supported by differential rotation is estimated to be about $1.5M_{max}$, so one finds $M_{rem}<1.5M_{max}$.  But the apparent existence of very high opacity heavy elements in the ejecta offers evidence against a long-lived uniformly-rotating supermassive remnant with $M_{max}<M_{rem}<M_{max,u}$.  The large neutrino flux from a surviving supermassive star would have protonized the ejecta and halted the nucleosynthesis of the heaviest elements.  Taking into account the ejected mass, binding energy corrections (more than 10\% of the rest masses), and the measured total gravitational mass of the components, $M_{tot}\simeq2.73-2.80M_\odot$, the condition $M_{rem}\gtrsim M_{max,u}$ suggests an upper limit to $M_{max}\lesssim2.2-2.3M_\odot$ \cite{margalit2017constraining}.

\begin{figure}
\centering\includegraphics[width=\linewidth,angle=0]{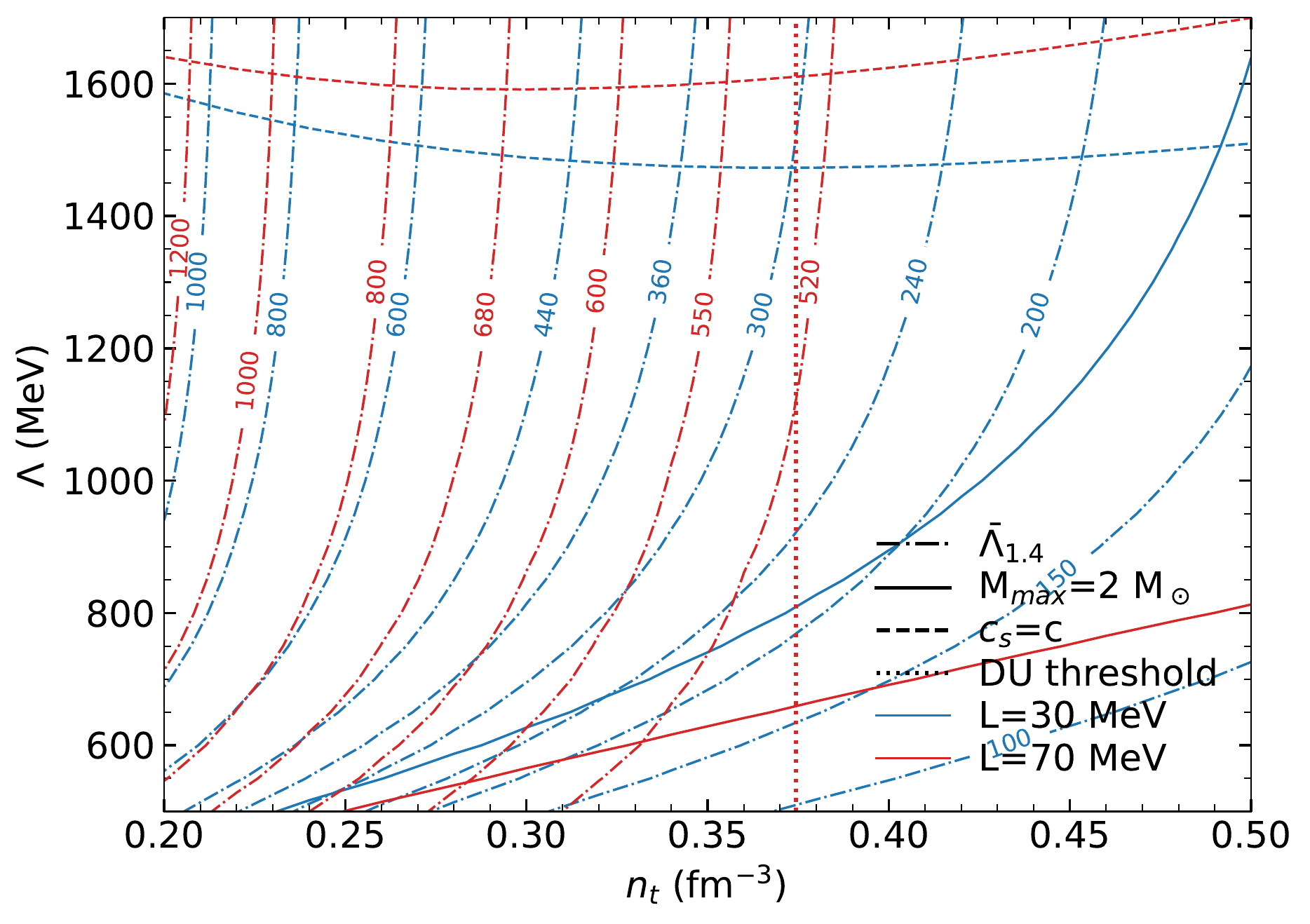}
\caption{The same as Fig. \ref{fig:p-space}, but showing tidal deformability $\bar\Lambda_{1.4}$ instead of $R_{1.4}$ contours.}
\label{fig:p-space-tidal14}\end{figure}
The mass, radius and causality constraints effectively bracket the ranges of allowed parameters, as can be seen in Fig. \ref{fig:p-space}.  The causality constraint effectively limits $\Lambda$ to values less than about 1600 MeV irresepective of assumed values for $n_t$ and $L$.  The constraint on $R_{1.4}$ largely limits $n_t$ to be larger than about 0.20 (0.25) fm$^{-3}$ for $\Lambda\gtrsim800$ MeV for $L=30 (70)$ MeV; smaller values of $n_t$ are allowed for $\Lambda\lesssim800$ MeV, but for $\Lambda\lesssim500$ MeV quarkyonic matter becomes possible only at unrealistically small values of $n_t\simeq n_0$.  The third boundary is set by the lower limit on the neutron star maximum mass; the greater is $M_{max}$, the more parameter space is confined.  

As an alternative to the radius constraint suggested by GW170817 and X-ray observations of neutron stars, GW170817 implies an upper limit to the tidal deformability of $1.4M_\odot$ neutron stars $\bar\Lambda_{1.4}<800$ \cite{abbott2017gw170817} or $\bar\Lambda_{1.4}\lesssim600$  \cite{de2018tidal, abbott2018gw170817}.  $\bar\Lambda$ is the dimensionless tidal deformability, effectively the constant of proportionality between  an external tidal field and the quadrupole deformation of a neutron star.  It can be straightforwardly determined from a first-order differential equation\citep{hinderer2010tidal,postnikov2010tidal} simultaneously integrated with the usual TOV differential equations for neutron star structure.  This deformability constraint is consistent, appproximately, with the condition $R_{1.4}\lesssim13$ km.  Parameter space constrained using $\bar\Lambda_{1.4}$ instead of $R_{1.4}$, together with $M_{max}>2M_\odot$ and causality, is shown in Fig. \ref{fig:p-space-tidal14}. 
\begin{figure*}
\includegraphics[width=0.495\linewidth,angle=0]{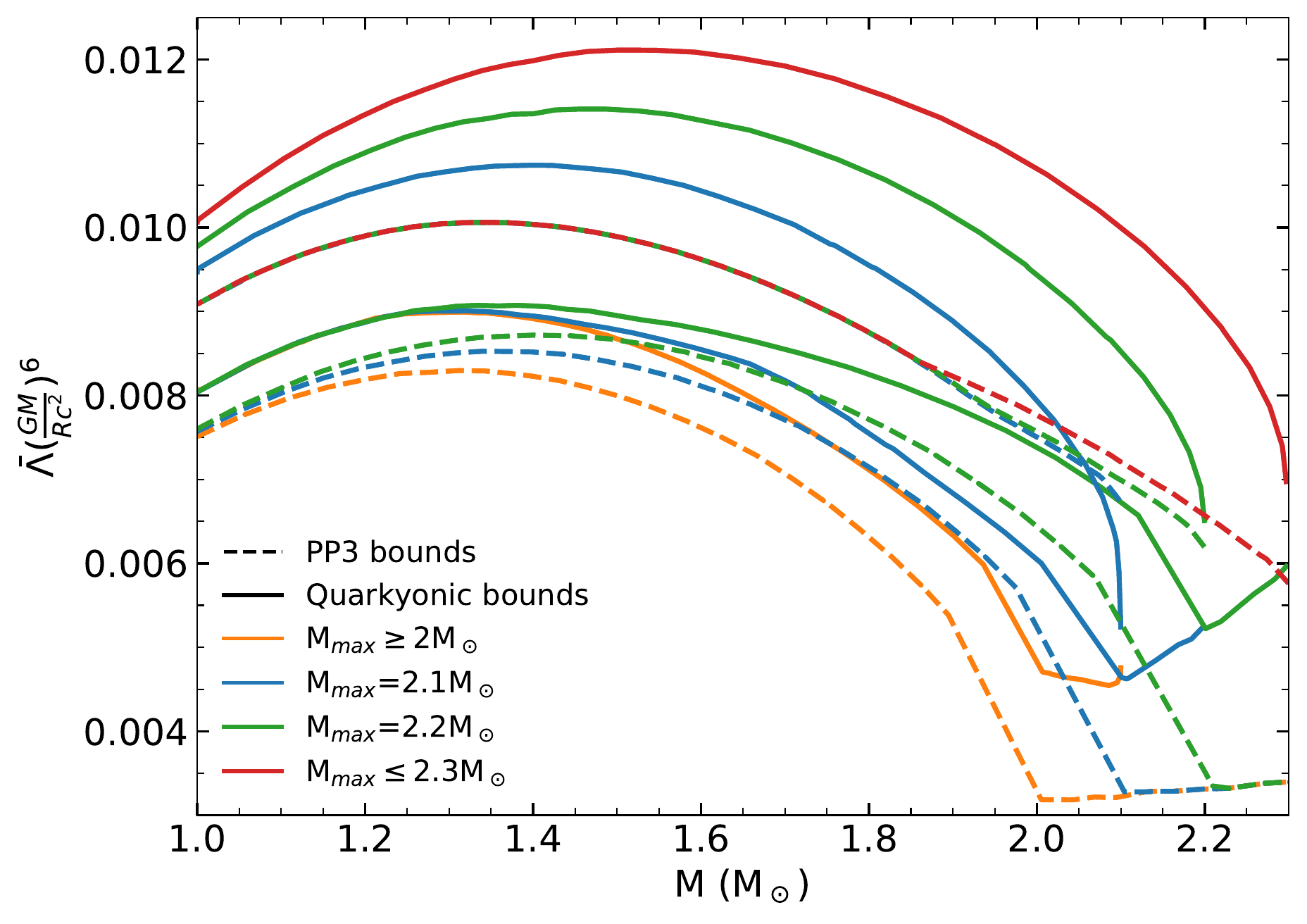}
\includegraphics[width=0.495\linewidth,angle=0]{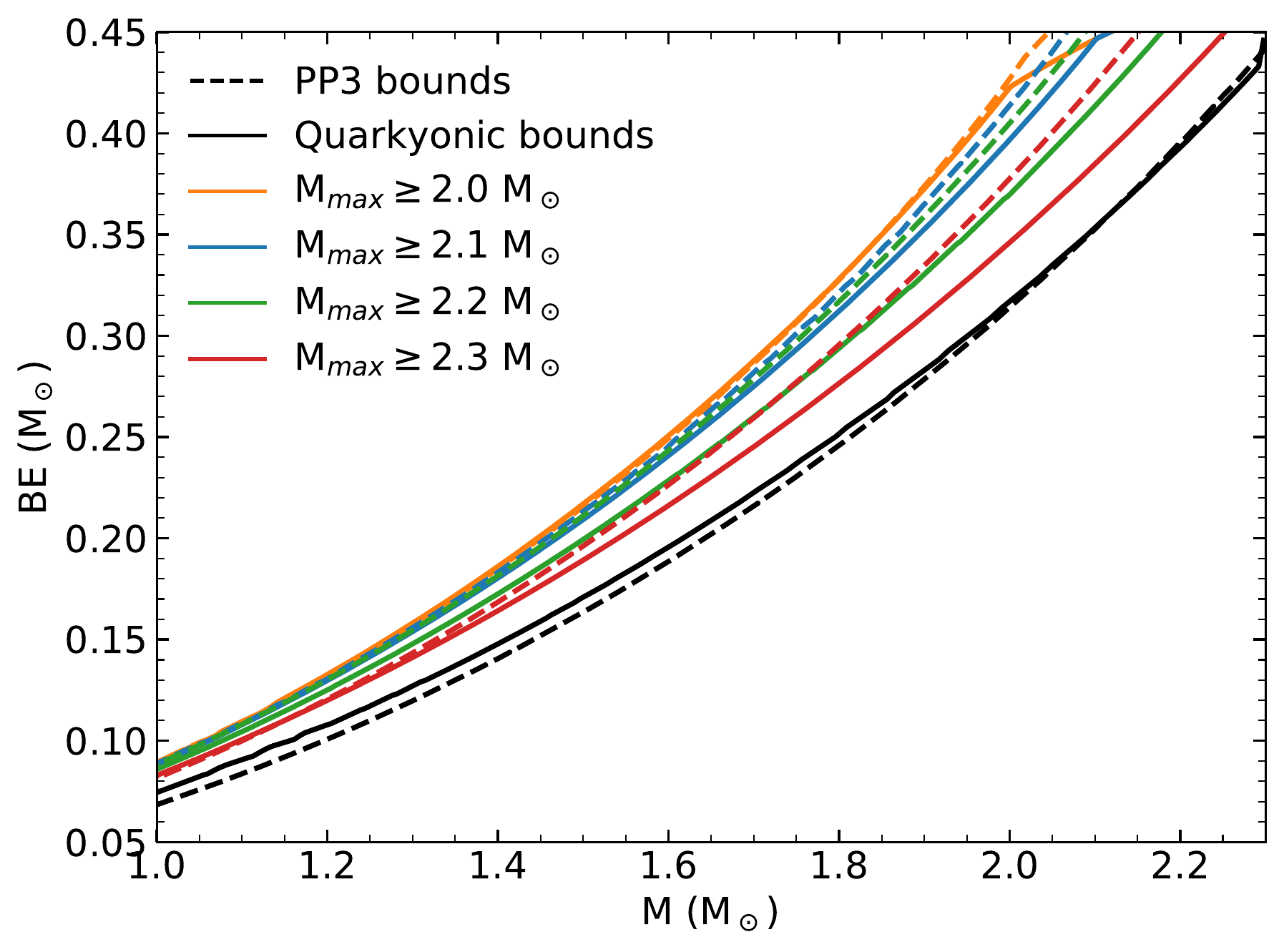}
\caption{Comparison of semi-universal relations for tidal deformability $\bar\Lambda$ and binding energy BE between quarkyonic (solid curves) and piecewise polytrope (dashed curves) parameterizations. The left (right) panel shows $\bar\Lambda(GM/Rc^2)^6$ (BE) as a function of mass.  In both cases, model parameters are constrained to satisfy 30 MeV $<L<70$ MeV, $R_{1.4}\le13.5$ km and selected maximum mass $M_{max}$ range.}
\label{fig:lam}\end{figure*}

Irrespective of the choice of constraints, Figs. \ref{fig:p-space} and \ref{fig:p-space-tidal14} indicate the ranges of permissible parameters are very large.  {\it Fine-tuning of the quarkyonic or hadronic model parameters, including the requirement that $n_t$ be very close to $n_s$ (for which there is no experimental support), is not needed  as is the situation for conventional hybrid quark-hadron models \citep{ranea2015constant,alford2013generic,han2019tidal,montana2019constraining}.} 

\section{Discussion and Conclusions\label{sec:disc}}

The modified quarkyonic EOS offers an alternative for the parameterized description of high-density matter in comparison to piecewise polytropes \cite{read2009constraints}, constant sound speed \cite{zdunik2013maximum,han2019tidal}, or spectral decomposition \cite{lindblom2010spectral} methods. Its rapidly varying sound speed and its narrow peak are features impossible to mimic with these approaches. The model presented here has only three parameters, making its parameterized use in statistical studies of observational data straightforward.  Nevertheless, additional parameters can be easily incorporated.  For example, one could replace the common $\Lambda$ with $\Lambda_n$ and $\Lambda_p$, allow d and u quarks to appear at different densities, and/or introduce more parameters to describe the hadronic phase.  

To demonstrate the utility of using the quarkyonic model as a parameterized high-density EOS, we compare the resulting bounds on some semi-universal relations for neutron stars with those established from three-parameter piecewise polytrope models (as used, for example, by \citet{ozel2009} and \citet{steiner2016}).  We will focus on relations involving the tidal deformability and binding energy.

\citet{zhao2018tidal} discovered, using piecewise polytropes, that $\bar\Lambda$  inversely correlates  with $(GM/Rc^2)^6$.  They found, for $1.1M_\odot<M<1.6M_\odot$ (the range of component masses inferred from the accurately determined chirp mass ${\cal M}=1.188M_\odot$ of GW170817), that the quantity $a=\bar\Lambda(GM/Rc^2)^6$ was confined to the relatively narrow range $a=0.0085\pm0.0010$ if  $M_{max}>2M_\odot$.    Fig. \ref{fig:lam} shows this correlation for quarkyonic model parameters restricted to 30 MeV $<L<70$ MeV and $\Lambda$ and $n_t$ bounded by the constraints $2M_\odot<M_{max}<2.3M_\odot$ and $R_{1.4}\le13.5$ km (see Fig. \ref{fig:p-space}).   In both parameterizations the lower bounds are slightly sensitive to the assumed minimum value of $M_{max}$.  The upper bounds, however, are quite sensitive to the assumed maximum value of $M_{max}$ for quarkyonic matter, while nearly independent of $M_{max}$ for piecewise polytropes.   The quarkyonic bounds are found to be $a=0.0099\pm0.0021$ with a somewhat larger range than determined using piecewise polytropes, $a=0.0089\pm0.0010$ \footnote{The fact that the range for piecewise polytropes is slightly different than that found in \cite{zhao2018tidal} is due to the constraint $R_{1.4}\le13.5$ km and 30 MeV $<L<70$ MeV imposed here.}.  This is not surprising, considering that the quarkyonic model samples more extreme pressure-energy density and sound speed-density behaviors.    In both cases, it is seen that restricting the range of $M_{max}$ reduces the uncertainty in $a$, but more so for quarkyonic stars than for hadronic stars.  Overall, it appears that piecewise polytropes may understate the uncertainty range for this correlation, an important consideration when attempting to deduce the EOS from observational data.

The binding energy is the difference between baryon and gravitational masses, BE = $M-M_{b}$, where $M_b$ is the total neutron star baryon number times the baryon rest mass.
Upper and lower bounds on this quantity as a function of $M$ are shown in Fig. \ref{fig:lam} and compared to analogous bounds derived from piecewise polytropes \citep{zhao2018tidal}.  In this case, there is less difference between the two approaches, and the lower bounds for each type of EOS are insensitive to the assumed $M_{max}$.  However, the upper bounds for each type of EOS decrease with assumed $M_{max}$, in contrast to the situation for $\bar\Lambda$.

The quarkyonic EOS also has implications for the cooling of neutron stars through the operation of the nucleon direct Urca (DU) process
\begin{eqnarray}
    p+e^- \rightarrow n + \nu_e,\qquad    n \rightarrow p +e^- + \bar \nu_e.
\label{eq:du}\end{eqnarray}
   The threshold for its operation is the kinematic condition 
   \begin{equation}
       k_{Fn}\le k_{Fp}+k_{Fe}, 
   \label{eq:kdu}     \end{equation}
        which requires a minimum $Y_L$ between 0.11 and 0.14 \citep{lattimer1991direct} within the hadronic sector.  This is never achieved for $L\lesssim50$ MeV for the particular interaction we employ as long as $n_B\lesssim0.7$ fm$^{-3}$.  Even in the stiffer cases, the threshold is not exceeded until relatively high densities are reached.  Because $Y_L$ abruptly decreases in quarkyonic matter for $n_B>n_t$,  it becomes increasingly hard to satisfy the kinematic constraint as the density increases in the quarkyonic sector.  Thus, 
the DU process can only operate if it is already permitted in hadronic matter at densities $n_B<n_t$.  Then it would operate in a shell centered at $n_t$, but would nonetheless be effective in rapidly cooling neutron stars.  According to the minimal cooling paradigm \cite{page2004minimal,page2013pairing}, most neutron star thermal emission observations are consistent with the lack of DU cooling.  The few exceptions could be explained by relatively massive neutron stars which might have central densities large enough for DU to operate.  This scenario would also fit quarkyonic stars, if $n_t$ is large enough and if $L$ is not too small.  Fig. \ref{fig:hbeta} explicitly shows the required conditions. 

Deconfined quarks can also participate in a direct Urca process
\begin{eqnarray}
d\rightarrow e^-+u+\bar\nu_e,\qquad u+e^-\rightarrow d+\nu_e,
\end{eqnarray}
having the kinematic requirement $k_{Fd}+k_{Fu}>k_{Fe}$.  This condition would be satisfied in quarkyonic matter at densities slightly in excess of $n_t$ irrespective of the value of $L$ because the quark abundances both grow rapidly with density.    However, the final momentum states of the quarks, which have to be above their Fermi surfaces, are blocked by nucleons occupying those states.  Therefore, a direct Urca process involving quarks may not be possible.

It is interesting to observe that quarkyonic configurations have the property that $R_{2.0}\ge R_{1.4}$, especially for large $\Lambda$ and small $n_t$ values. This results from a positive slope $dR/dM$ at moderate masses. $(c^2/G)(dR/dM)_{1.4}\gtrsim1$ can always be achieved for $1.4M_\odot$ stars as long as $n_t$ is small and $\Lambda$ is large, no matter how soft the symmetry energy is. For example, hadronic (quarkyonic) stars with $L=30, 50$ or 70 MeV have  $(c^2/G)(dR/dM)_{1.4}=-1.25(1.16), -0.48(1.09)$ or $-0.5(1.0)$, respectively, assuming $n_t=0.20$ fm$^{-3}$ and $\Lambda=1600$ MeV for the quarkyonic stars.   This may be of interest in view of the forthcoming NICER radius measurements of PSR J0740+6620 whose mass is estimated to be $2.14^{+0.10}_{-0.09}M_\odot$ from pulsar timing observations \citep{cromartie2020relativistic}.  The large measured mass of PSR J0740+6620 contrasts with the lower mass estimate associated with another NICER target, PSR J0030+0451 with $1.44^{+0.15}_{-0.14}M_\odot$ \citep{raaijmakers2019nicer,miller2019psr}.  A measurement resulting in evidence that $R_{2.0}\gtrsim R_{1.4}$ would bolster support for a quarkyonic-like dense matter EOS with a sharp peak in $c_s$ in the vicinity of $2-4 n_s$.

In summary, we have formulated a model of quarkyonic matter that can be a useful tool for parameterizing the high-density EOS.  Parameterized high-density EOSs are frequently used in the interpretation of astrophysical observations.  The quarkyonic model has several advantages relative to alternative models involving piecewise polytropes, power-law expansions, relativistic mean-field theory or spectral decomposition. It can be used to simulate hybrid stars satisfying observed maximum mass and radius constraints without forcing the expected quark-hadron transition to lie abnormally close to the nuclear saturation density. { In fact, if one desires a simpler high-density parameterized EOS with all the advantages and physical motivations of the beta-equilibrium quarkyonic model but without concern for its compositional details, we develop in Appendix \ref{app:simple} a pure-neutron matter version, also with three parameters, that is completely analytic and therefore particularly convenient for the interpretation of observational data.}   

{
\appendix
\section{Simplified $ndu$ Quarkyonic Model}\label{app:simple}

In many applications of parameterized EOSs, it is not necessary to consider the underlying compositional details, such as the presence of leptons or beta equilibrium.  Here, we present a simpler version of quarkyonic matter involving chemical equilibrium among neutrons and d and u quarks that has the advantageous sound-speed behavior of the beta-equilibrium version and is also completely analytic.  Because in beta equilibrium the lepton fractions are generally small, this $ndu$ version closely mimics the behavior of the beta-equilibrium version.

Nucleons are described as in the beta-equilibrium version.  For $n_B<0.5n_s$ we employ the SLy4 crust EOS. For greater densities, we assume the nucleon potential of Eq. (\ref{eq:pot}) with $n_p=0$, that is
\begin{equation}
 {    V(n_n)=a_1u+b_1u^{\gamma_1}},
     \end{equation}
where $u=n_n/n_s$. For densities between the crust and $n_t$ the baryon density and chemical potential are
\begin{eqnarray}
 n_B&=&n_n=k_{Fn}^3/(3\pi^2),\\
 \mu&=&\mu_n=\sqrt{m_B^2+(\hbar k_{Fn}c)^2}+{\partial n_nV(n_n)\over\partial n_n}.\label{eq:mun}
 \end{eqnarray}
$\varepsilon=\varepsilon_B$ is given by Eq. (\ref{eq:enuc}) for neutrons with $k_{0n}=0$.

 Quarkyonic matter appears at the density $n_{t}$ and is in chemical equilibrium such that
 \begin{equation}
     \mu_n=2\mu_d+\mu_u.
     \label{eq:chem}
     \end{equation}
 The properties of $ndu$ quarkyonic matter are insensitive to the mass ratio $m_d/m_u$ as long as it is of order of magnitude unity; therefore, we assume $m_u=m_d=\mu_{nt}/3$. 

In the quarkyonic sector,  charge neutrality requires 
    $k_{Fu}^3=k_{Fd}^3/2$. 
Eq. (\ref{eq:chem}) has the analytic solution for $k_{Fd}$ as a function of $k_{Fn}$:
\begin{equation}
    k_{Fd}={\mu_n\over\hbar cC}\left(8-C-3C{m_d^2/\mu_n^2}-4\sqrt{Q}\right)^{1/2},\label{eq:kfd}
\end{equation}
where $C=4-2^{-2/3}$ and $Q=2^{-2/3}+C(C-3)m_d^2/\mu_n^2$. Eq. (\ref{eq:chempo_neutron}) gives $\mu_n(k_{Fn})$ where $k_{0n}$ is given by Eq. (\ref{eq:k0}).  The total baryon density is
\begin{equation}
    n_B={k_{Fn}^3-k_{0n}^3\over3\pi^2}+{k_{Fd}^3\over2\pi^2},
\end{equation}
{and is a monotonically increasing function of $k_{Fn}$ for all parameter values}. The neutron and quark energy densities are given by Eqs. (\ref{eq:enuc}) and (\ref{eq:nif}) {as in the full model;} the neutron wavenumber saturates for $n_B>n_t$ to the value $k_{mn}$ according to Eq. (\ref{eq:kmax}){, as long as $\Lambda/(\hbar k_{tn})>0$}.  

\begin{figure}[ht]
\includegraphics[width=\linewidth,angle=0]{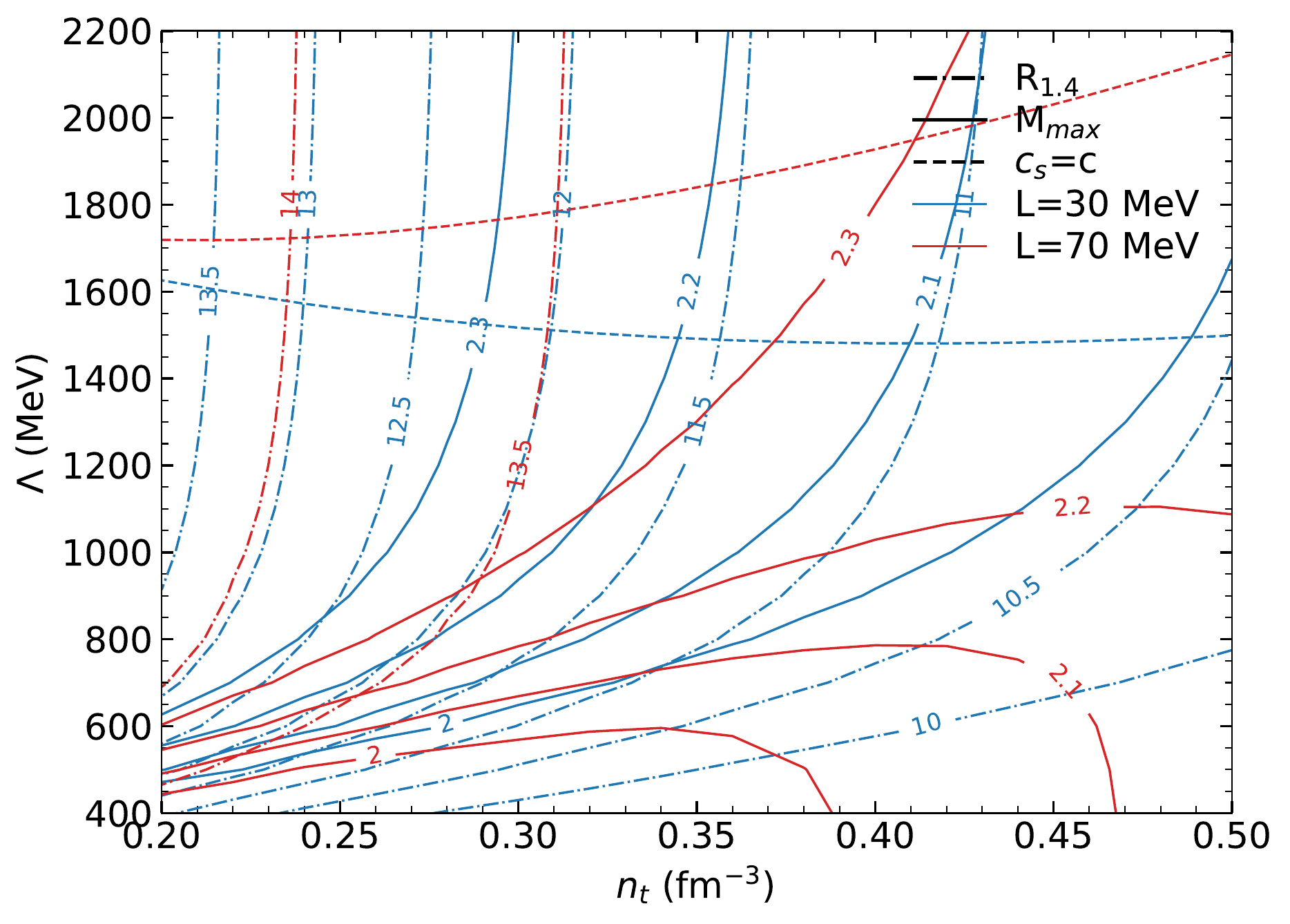}
\caption{Allowed parameter space for the $ndu$ version of the quarkyonic matter EOS. Similar to Fig. \ref{fig:p-space} showing contours of $R_{1.4}$, $M_{max}$ and $c_s^2=1$ for $L=30$ MeV and $L=70$ MeV.}
\label{fig:p-space-PNM}\end{figure}

For given values of $n_t$, $\Lambda$ and $L$, this model gives very similar results as for the full model.  For comparison, Fig. \ref{fig:p-space-PNM} shows the allowed parameter space of the $ndu$ version of the quarkyonic matter EOS. 
}

\vspace*{0.1cm}
\section*{Acknowledgements}
We thank S. Reddy and M. Prakash for helpful comments.  This work was supported in part by NASA through the NICER mission with  NASA Grant 80NSSC17K0554 and by the U.S. DOE from Grant DE-FG02-87ER40317.

\bibliography{quarkyonic}

\end{document}